\documentclass[superscriptaddress,showpacs,amssymb,amsmath,
amsfonts,aps,
prl,twocolumn
]{revtex4}
\usepackage{graphicx}
\usepackage{dcolumn}
\usepackage{epsfig}
\usepackage{subfigure}
\begin{document}

\title{Branching Ratio of the Electromagnetic Decay of the $\Sigma^{+}(1385)$}

\newcommand{\bra}[1]{\left\langle #1 \right|}
\newcommand{\ket}[1]{\left| #1 \right\rangle}
\newcommand{\bracket}[2]{\left\langle #1 | #2 \right\rangle}
\newcommand{\matrixelement}[3]{\bra{#1} \hat{#2} \ket{#3}}
\newcommand\polvec{\vec\epsilon_\lambda^{\,*}(\vec k)}
\newcommand\etal{{\em et al.}}

\pacs{13.40.Em,14.20.Jn,13.30.Ce,13.40.Hq}

\newcommand*{\ANL}{Argonne National Laboratory, Argonne, Illinois 60439}
\newcommand*{\ANLindex}{1}
\affiliation{\ANL}
\newcommand*{\ASU}{Arizona State University, Tempe, Arizona 85287-1504}
\newcommand*{\ASUindex}{2}
\affiliation{\ASU}
\newcommand*{\CANISIUS}{Canisius College, Buffalo, NY}
\newcommand*{\CANISIUSindex}{3}
\affiliation{\CANISIUS}
\newcommand*{\CMU}{Carnegie Mellon University, Pittsburgh, Pennsylvania 15213}
\newcommand*{\CMUindex}{4}
\affiliation{\CMU}
\newcommand*{\CUA}{Catholic University of America, Washington, D.C. 20064}
\newcommand*{\CUAindex}{5}
\affiliation{\CUA}
\newcommand*{\SACLAY}{CEA, Centre de Saclay, Irfu/Service de Physique Nucl\'eaire, 91191 Gif-sur-Yvette, France}
\newcommand*{\SACLAYindex}{6}
\affiliation{\SACLAY}
\newcommand*{\CNU}{Christopher Newport University, Newport News, Virginia 23606}
\newcommand*{\CNUindex}{7}
\affiliation{\CNU}
\newcommand*{\UCONN}{University of Connecticut, Storrs, Connecticut 06269}
\newcommand*{\UCONNindex}{8}
\affiliation{\UCONN}
\newcommand*{\EDINBURGH}{Edinburgh University, Edinburgh EH9 3JZ, United Kingdom}
\newcommand*{\EDINBURGHindex}{9}
\affiliation{\EDINBURGH}
\newcommand*{\FU}{Fairfield University, Fairfield CT 06824}
\newcommand*{\FUindex}{10}
\affiliation{\FU}
\newcommand*{\FIU}{Florida International University, Miami, Florida 33199}
\newcommand*{\FIUindex}{11}
\affiliation{\FIU}
\newcommand*{\FSU}{Florida State University, Tallahassee, Florida 32306}
\newcommand*{\FSUindex}{12}
\affiliation{\FSU}
\newcommand*{\Genova}{Universit$\grave{a}$ di Genova, 16146 Genova, Italy}
\newcommand*{\Genovaindex}{13}
\affiliation{\Genova}
\newcommand*{\GWUI}{The George Washington University, Washington, DC 20052}
\newcommand*{\GWUIindex}{14}
\affiliation{\GWUI}
\newcommand*{\ISU}{Idaho State University, Pocatello, Idaho 83209}
\newcommand*{\ISUindex}{15}
\affiliation{\ISU}
\newcommand*{\INFNFE}{INFN, Sezione di Ferrara, 44100 Ferrara, Italy}
\newcommand*{\INFNFEindex}{16}
\affiliation{\INFNFE}
\newcommand*{\INFNFR}{INFN, Laboratori Nazionali di Frascati, 00044 Frascati, Italy}
\newcommand*{\INFNFRindex}{17}
\affiliation{\INFNFR}
\newcommand*{\INFNGE}{INFN, Sezione di Genova, 16146 Genova, Italy}
\newcommand*{\INFNGEindex}{18}
\affiliation{\INFNGE}
\newcommand*{\INFNRO}{INFN, Sezione di Roma Tor Vergata, 00133 Rome, Italy}
\newcommand*{\INFNROindex}{19}
\affiliation{\INFNRO}
\newcommand*{\ORSAY}{Institut de Physique Nucl\'eaire ORSAY, Orsay, France}
\newcommand*{\ORSAYindex}{20}
\affiliation{\ORSAY}
\newcommand*{\ITEP}{Institute of Theoretical and Experimental Physics, Moscow, 117259, Russia}
\newcommand*{\ITEPindex}{21}
\affiliation{\ITEP}
\newcommand*{\JMU}{James Madison University, Harrisonburg, Virginia 22807}
\newcommand*{\JMUindex}{22}
\affiliation{\JMU}
\newcommand*{\KNU}{Kyungpook National University, Daegu 702-701, Republic of Korea}
\newcommand*{\KNUindex}{23}
\affiliation{\KNU}
\newcommand*{\LPSC}{LPSC, Universite Joseph Fourier, CNRS/IN2P3, INPG, Grenoble, France
}
\newcommand*{\LPSCindex}{24}
\affiliation{\LPSC}
\newcommand*{\UNH}{University of New Hampshire, Durham, New Hampshire 03824-3568}
\newcommand*{\UNHindex}{25}
\affiliation{\UNH}
\newcommand*{\NSU}{Norfolk State University, Norfolk, Virginia 23504}
\newcommand*{\NSUindex}{26}
\affiliation{\NSU}
\newcommand*{\OHIOU}{Ohio University, Athens, Ohio  45701}
\newcommand*{\OHIOUindex}{27}
\affiliation{\OHIOU}
\newcommand*{\ODU}{Old Dominion University, Norfolk, Virginia 23529}
\newcommand*{\ODUindex}{28}
\affiliation{\ODU}
\newcommand*{\RPI}{Rensselaer Polytechnic Institute, Troy, New York 12180-3590}
\newcommand*{\RPIindex}{29}
\affiliation{\RPI}
\newcommand*{\URICH}{University of Richmond, Richmond, Virginia 23173}
\newcommand*{\URICHindex}{30}
\affiliation{\URICH}
\newcommand*{\ROMAII}{Universita' di Roma Tor Vergata, 00133 Rome Italy}
\newcommand*{\ROMAIIindex}{31}
\affiliation{\ROMAII}
\newcommand*{\MSU}{Skobeltsyn Nuclear Physics Institute, Skobeltsyn Nuclear Physics Institute, 119899 Moscow, Russia}
\newcommand*{\MSUindex}{32}
\affiliation{\MSU}
\newcommand*{\SCAROLINA}{University of South Carolina, Columbia, South Carolina 29208}
\newcommand*{\SCAROLINAindex}{33}
\affiliation{\SCAROLINA}
\newcommand*{\JLAB}{Thomas Jefferson National Accelerator Facility, Newport News, Virginia 23606}
\newcommand*{\JLABindex}{34}
\affiliation{\JLAB}
\newcommand*{\UNIONC}{Union College, Schenectady, NY 12308}
\newcommand*{\UNIONCindex}{35}
\affiliation{\UNIONC}
\newcommand*{\UTFSM}{Universidad T\'{e}cnica Federico Santa Mar\'{i}a, Casilla 110-V Valpara\'{i}so, Chile}
\newcommand*{\UTFSMindex}{36}
\affiliation{\UTFSM}
\newcommand*{\GLASGOW}{University of Glasgow, Glasgow G12 8QQ, United Kingdom}
\newcommand*{\GLASGOWindex}{37}
\affiliation{\GLASGOW}
\newcommand*{\VIRGINIA}{University of Virginia, Charlottesville, Virginia 22901}
\newcommand*{\VIRGINIAindex}{38}
\affiliation{\VIRGINIA}
\newcommand*{\WM}{College of William and Mary, Williamsburg, Virginia 23187-8795}
\newcommand*{\WMindex}{39}
\affiliation{\WM}
\newcommand*{\YEREVAN}{Yerevan Physics Institute, 375036 Yerevan, Armenia}
\newcommand*{\YEREVANindex}{40}
\affiliation{\YEREVAN}

\newcommand*{\NOWLANL}{Los Alamos National Laboratory, Los Alamos, NM 87544 USA}
\newcommand*{\NOWMSU}{Skobeltsyn Nuclear Physics Institute, Skobeltsyn Nuclear Physics Institute, 119899 Moscow, Russia}
\newcommand*{\NOWINFNFR}{INFN, Laboratori Nazionali di Frascati, 00044 Frascati, Italy}
\newcommand*{\NOWINFNGE}{INFN, Sezione di Genova, 16146 Genova, Italy}
\newcommand*{\NOWANL}{Argonne National Laboratory, Argonne, Illinois 60439}

\author {D.~Keller} 
\affiliation{\OHIOU}
\author {K.~Hicks} 
\affiliation{\OHIOU}
\author {K.P. ~Adhikari} 
\affiliation{\ODU}
\author {D.~Adikaram} 
\affiliation{\ODU}
\author {M.J.~Amaryan} 
\affiliation{\ODU}
\author {M.~Anghinolfi} 
\affiliation{\INFNGE}
\author {H.~Baghdasaryan} 
\affiliation{\VIRGINIA}
\affiliation{\ODU}
\author {J.~Ball} 
\affiliation{\SACLAY}
\author {M.~Battaglieri} 
\affiliation{\INFNGE}
\author {I.~Bedlinskiy} 
\affiliation{\ITEP}
% ...

\collaboration{The CLAS Collaboration}
\noaffiliation
\date{\today}

\begin{abstract}
The CLAS detector was used to obtain the first ever measurement of the electromagnetic decay
of the $\Sigma^{*+}(1385)$ from the reaction $\gamma p \to K^0 \Sigma^{*+}(1385)$.
A real photon beam with a maximum energy of 3.8 GeV
was incident on a liquid-hydrogen target, resulting in the
photoproduction of the kaon and $\Sigma^*$ hyperon.  Kinematic fitting was used to separate
the reaction channel from the background processes.  The fitting algorithm exploited a new method to kinematically fit neutrons in the CLAS detector, leading
to the partial width measurement of $250.0\pm56.9(stat)^{+34.3}_{-41.2}(sys)$ keV.  A U-spin symmetry test using the SU(3) flavor-multiplet representation yields predictions for the $\Sigma^{*+}(1385)\to\Sigma^{+}\gamma$ and $\Sigma^{*0}(1385)\to\Lambda\gamma$ partial widths that agree with the experimental measurements. 
\end{abstract}

\maketitle

\section{Introduction}

The electromagnetic (EM) decay of baryons can provide considerable information on their underlying
structure.  This transition offers a clean
probe of the wavefunction of the initial and final state baryons, providing theoretical constraints
and tests of the quark model.  The non-relativistic quark model (NRQM) of Isgur
and Karl \cite{Isgur1,Isgur2} predicts the electromagnetic properties of the ground state baryons reasonably
well.  It has been less successful giving  accurate descriptions of the low-lying excited-state
hyperons.  Several other theoretical techniques have been used to more accurately 
calculate these transitions, including 
(NRQM) \cite{DHK,Koniuk}, a relativized constituent quark model (RCQM) \cite{warns}, a chiral
constituent quark model ($\chi$CQM) \cite{wagner}, the MIT bag model \cite{kaxiras}, 
the bound-state soliton model \cite{Schat}, a three-flavor generalization of the Skyrme model
that uses the collective approach \cite{Abada,Haberichter}, and an algebraic model of hadron structure
\cite{Bijker}.

Photoproduction from nucleon targets is a useful technique to cleanly
generate a significant statistical sample of hyperons and to measure EM
transitions to other decuplet baryons.
If the EM transition form factors for 
decuplet baryons with strangeness are also sensitive to meson 
cloud effects, models attempting to make predictions of the decuplet radiative decay widths
will need revisions to incorporate this effect.  Comparison of data for 
the EM decay of decuplet hyperons, $\Sigma^*$, to the present predictions 
of quark models provides a measure of the importance of meson cloud 
diagrams in the $\Sigma^* \to Y\gamma$ transition.  Experimental
results for the EM decay ratios for all $\Sigma^*$ charge states
are desirable to obtain a complete comparison to EM decay predictions
for the $\Sigma^*$.  Precision measurements of the $\Sigma^{*-}\to\Sigma^-\gamma$ and $\Sigma^{*+}\to\Sigma^+\gamma$ decay widths can be particularly useful in determining
the degree of SU(3) symmetry breaking.

The decay width from the measurement of $\Sigma^{*0}\to \Lambda \gamma$ \cite{kellpaper,taylor} is
much larger than most current theoretical predictions.  This could be due
to meson cloud effects, which were not included in these calculations.  There is a theoretical basis for calculating these effects \cite{Lee} that suggests pion cloud
effects may be sizable.  For example, they are predicted to contribute on the order of $\sim$40\% to the $\gamma p \to \Delta^+$  
magnetic dipole transition form factor, $G^\Delta_M(Q^2)$, for low $Q^2$.
The CQM \cite{CQM} indicates that the value of $G^\Delta_M(0)$ is directly proportional to 
the proton magnetic moment \cite{Lee}, and measurements of $G^\Delta_M$ 
for low $Q^2$ are rationalized in the framework of the model 
if the experimental magnetic moment is lowered by about 25\%. 

With theoretical predictions for the degree at which the meson cloud effects
plays a role, it is then possible to test $SU(3)$ flavor symmetry breaking
(and the degree at which it is broken).  This can be achieved by measuring both the
$\Gamma(\Sigma^{*-}\to\Sigma^-\gamma)$ and $\Gamma(\Sigma^{*+}\to\Sigma^+\gamma)$ decay widths and comparing these to predictions from flavor SU(3) relations.

Just like isospin invariance can be used to compare the $\Delta^{++} \to
p \pi^+$ and $\Delta^+ \to p \pi^0$ decays, U-spin invariance may be used to
compare the $\Sigma^{*+} \to \Sigma^+ \gamma$ and $\Delta^+ \to p \gamma$ 
decays. U-spin is analogous to isospin in
that it is a symmetry in the exchange of the $d$ and $s$ quarks rather than
the $u$ and $d$ quarks.  A value of U-spin can be assigned to each baryon
based on its quark composition.  The $\Sigma^{*-}$ and the $\Xi^{*-}$ of
the baryon decuplet have $U=3/2$, whereas the octet baryons $\Sigma^-$ and
$\Xi^-$ have $U=1/2$.

U-spin symmetry forbids radiative decays 
of specific decuplet baryons.  Since the photon 
is a charge singlet with $U=0$, this implies that
$$ \Sigma^{*-} \to \Sigma^- \gamma \  {\rm and} \ 
   \Xi^{*-} \to \Xi^- \gamma $$
have zero amplitude in the equal-mass limit due to U-spin symmetry.
This can also be understood in the context of the SU(6) wavefunctions for 
these baryons.  The M1 transition operator is written between the initial 
and final states as :
\begin{equation}
\langle \Sigma^-_{SU(6)} | \, \sum_q \frac{Q_q}{2m_q}\sigma_q
\cdot \left( \bf k \rm \times \epsilon^{* \lambda}\right) \, | 
   \Sigma^{*-}_{SU(6)} \rangle = 0.
\label{eq1}
\end{equation}
Here the sum is over all $q$ constituent quarks, $m_q$, $\sigma_q$ and $Q_q$ are the mass, spin vector and charge of the $q$ quark, $\bf k \rm$ is the propagation direction, and $\epsilon^{* \lambda}$ is the polarization vector. 
One can also show that the same transition operator for the $\Sigma^{*+}$ gives a 
non-zero amplitude.
U-spin invariance implies a large difference in the radiative decay widths of the 
$\Sigma^{*-}$ and $\Sigma^{*+}$.

The chiral symmetry for U-spin is strongly broken because the {\it constituent} mass of the strange quark, $m_s$, is 
approximately 1.5 times greater than the non-strange quarks, $m$. 
The magnetic moment is inversely proportional to the mass, 
and so there is no cancellation in the wavefunction like in the equal-mass $SU(6)$ 
case in Eq. \ref{eq1}. From Ref. \cite{lipkin2}, an estimate 
of the ratio of the EM decay rates from the ratio of the square of the 
transition operators can be expressed as
$$ \frac{ \Gamma ( \Sigma^{*-} \to \Sigma^- \gamma ) }
        { \Gamma ( \Sigma^{*+} \to \Sigma^+ \gamma ) } \approx
   \frac{1}{9} \left( 1 - \frac{m}{m_s} \right)^2, $$ resulting in a value of about 1\%. This suggests that U-spin 
symmetry breaking for radiative decays is at the level of only a few percent.
At this level U-spin is an effective tool, even considering 
the quark mass difference.

Detailed calculations from the CQM and $1/N_c$-type expansions of the EM decay rates have been carried out by several  
groups \cite{quarks, nrqm}, all of which come up with decay
ratios of a similar scale.  In lattice QCD, the quarks have very different interactions with 
the photon than for the CQM, but these too have ratios (for the above equation) within a few percent \cite{lattice}.
This consistency makes a stronger case for the usefulness of U-spin symmetry. 

There has been much theoretical interest in radiative baryon 
decays. However, there are only a few measurements.  Recently, a measurement of
the radiative decay of the $\Sigma^{*-}$ was attempted 
by the SELEX collaboration \cite{selex}, resulting in only 
an upper limit.  The 90\% confidence level upper bound of 
$\Gamma=9.5$ keV was reported, however most models predict a 
value of less than 4 keV.  Ultimately this result has limited power to constrain 
theoretical estimates.  More experimental measurements
are necessary to provide better constraints.

A program to investigate the various $\Sigma^{*}$
electromagnetic decays is underway using data from the CEBAF Large Angle Spectrometer (CLAS) detector.  First, two independent analyses
of the EM decay of the $\Sigma^{*0}$ have been completed \cite{kellpaper}, \cite{taylor}.  The consistency in 
these results has given confidence in the notion that meson cloud effects are indeed contributing
significantly.  The next step described and presented here was to measure the $\Sigma^{*+}$ electromagnetic decay, which
has not been done before.  The final program analysis for the $\Sigma^{*-}\to\Sigma^-\gamma$ decay will be addressed in a future CLAS publication.

In the following, a description of the experimental details and analysis procedure for
extracting the $\Sigma^{*+}$ EM decay branching ratio normalized to the strong decay is provided.  
Some specifics are given about neutron detection and the development of the
neutron covariance matrix required by the analysis.  After the signal extraction a U-spin
symmetry test using the U-spin SU(3) multiplet representation is used to
predict the $\Sigma^{*+}\to\Sigma^{+}\gamma$ and $\Sigma^{*0}\to\Lambda\gamma$
partial widths, which are then compared to the experimental results.

\section{The Experiment}

The present measurements were carried out with the CLAS in Hall B at the Thomas Jefferson National
Accelerator Facility \cite{mecking}.  An electron beam of energy 4.023 GeV was
used to produce a photon beam with an energy range of 1.6-3.8 GeV, as
deduced by a magnetic spectrometer \cite{tagnim} that ``tagged" the electron
with an energy resolution of $\sim0.1\%$.  A 40-cm-long liquid hydrogen
target was placed such that the center of the target was 10 cm
upstream from the center of CLAS.

\begin{figure}
\epsfig{file=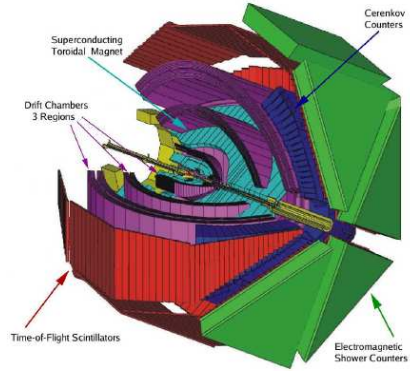,width=\columnwidth}
\caption{The CLAS detector at Jefferson Lab showing the toroidal magnet,
the drift chambers, the time-of-flight scintillators, the cerenkov counter,
and the electromagnetic calorimeter.}
\label{fullclas}
\end{figure}

The CLAS detector is 
constructed around six superconducting coils that generate
a toroidal magnetic field to momentum-analyze charged
particles. The detection system consists of multiple layers
of drift chambers to determine charged-particle trajectories, Cerenkov detectors for electron/pion separation, scintillation counters for flight-time measurements,
and calorimeters to identify electrons and high-energy neutral particles, see Fig. \ref{fullclas}.
The Cerenkov detectors are not required for this experiment.

Each event trigger required a coincidence between the OR of the
detector elements in the focal plane of the photon spectrometer and the CLAS Level 1 trigger.
The Level 1 trigger required two charged particles in two different sectors
of CLAS within a 150 ns coincidence time window.  The approximate integrated luminosity
for the CLAS $g11a$ run period used in this analysis was 70 pb$^{-1}$.
Details of the experimental setup can be found in Refs.  
\cite{mecking,mccracken}.

\section{Event Selection}
Events were selected for the channel $\gamma
p \to K^{0} \Sigma^{*+}$.  The present Particle Data Group (PDG) branching ratios list the decay $\Sigma^{*} \to \Sigma\pi$ to be $11.7\pm1.5\%$, and assuming isospin symmetry, this
leads to a branching ratio of $5.85\pm0.75\%$ for the $\Sigma^{*+} \to
\Sigma^{+}\pi^{0}$ decay \cite{pdg}.  This channel will be used to normalize the radiative
signal that comes from the channel $\Sigma^{*+} \to \Sigma^{+}\gamma$.  For both
channels the topology of the decay is $\gamma p \to K^{0} \Sigma^{+}(X)$,
where $X$ is not directly measured, such that the $\pi^{0}$ and $\gamma$ are
differentiated using conservation of energy and momentum.  This topology leads to the final set of decay products 
$\gamma p \to K^{0} \Sigma^{+}(X)\to \pi^{+}\pi^{-}\pi^{+}n(X)$.  The charged
particles can easily be detected with the use of the CLAS drift chambers and
time-of-flight system.  The neutron must be detected with the CLAS electromagnetic
calorimeters.  The analysis was done using a previously prepared data reduction (skim) that required
two positively charged tracks and one negatively charged track for each event. 

Cuts were applied to take into account both the regions of CLAS where there are holes in the acceptance that arose from problematic detector elements and regions that were not well simulated.
This includes tracks at extremely forward or backward angles, areas near the torus coils, and regions where the drift chambers and scintillator counter efficiencies
were not well understood.  Tracks that point near these shadow regions are less likely to be reconstructed accurately.  In addition a minimum momentum of 0.125 GeV, after energy loss corrections, was enforced for both positively and
negatively charged particles to ensure accurate drift chamber track reconstruction.

During the initial data skim, the hit times in the start counter that surround the target were used to find an interaction vertex time for each charged particle, which was then matched
up with photons identified in the tagger, where there can be up to 10 candidate photons for a given event.  The photon with the closest time to any track was selected as the photon that caused the event.
Specifically, the time of interaction was determined using the time of the electron beam bucket (the accelerator RF time) that produced the event.  To
correlate the interaction time with the photon production time, a timing coincidence between the tagger and the start counter was used. The  RF time for the photon was then used to get the vertex time (photon interaction time $t_{\gamma}$) for the event.  Using the time-of-flight (TOF) from
the event vertex to the scintillator counter, the velocity $\beta$ was
calculated for each particle.  From $\beta$ and the particle's measured momentum, a mass was calculated.  Each track did not need to have a hit registered in the start counter for its mass to be calculated, only one track in the event needed a start counter hit. 

The mass squared calculated from time-of-flight is
\begin{equation}
m^{2}_{cal} = \frac{p^{2}(1 - \beta^{2})}{\beta^{2}},
\end{equation}
where $\beta = L/t_{meas}$ such that $L$ is the path length from the target to the scintillator, $t_{meas}$ is the measured
time-of-flight, and the speed of light is set to 1.
From this initial identification, it was possible to use additional timing information to improve event
selection.  The measured time-of-flight and calculated time-of-flight were used for an additional constraint.  The measured time-of-flight is $t_{meas} = t_{sc}-t_{\gamma}$, where $t_{sc}$ is the time at which the particle strikes the time-of-flight scintillator counter.  $\Delta t$ is then
\begin{equation}
\Delta t = t_{meas} - t_{cal},
\end{equation}
where $t_{cal}$ is the time-of-flight calculated for an assumed mass such that
\begin{equation}
t_{cal} = L\sqrt{1 + { \left(m \over p \right)}^{2}},
\end{equation}
where $m$ is the assumed mass for the particle of interest, and $p$ is the
momentum magnitude.  Cutting on $\Delta t$ or $m_{cal}$ should be effectively equivalent.

Using $\Delta t$ for each particle it was possible to reject events that were not associated with the correct RF beam bucket, which was
separated by 2 ns.  This was done by requiring $|\Delta t| \leq$1 ns for all charged particles in the initial analysis.  This cut was chosen to minimize signal loss while also minimizing overlap from
other beam buckets. 

\begin{figure}
\epsfig{file=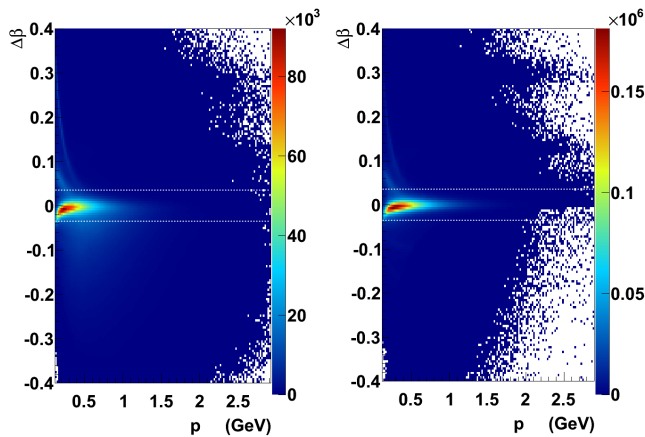,width=\columnwidth}
\caption{The $\Delta \beta$ distributions for $\pi^+$ (left) and $\pi^-$ (right).  The
cut of $-0.035\leq \Delta \beta \leq 0.035$ is shown as the dashed lined in each case.}
\label{dbeta}
\end{figure}

A $\Delta \beta$ cut was used to clean up the identification scheme.  $\Delta \beta$ is the difference between the measured $\beta=L/(t_{meas})$ and the calculated $\beta =
p/\sqrt{p^2+m^2}$.  The good events were taken within a cut of $-0.035\leq \Delta \beta \leq 0.035$ for all pions as shown in Fig. \ref{dbeta}.
\subsection{Kaon identification}
\label{kid}
In the reaction of interest, $\gamma p \to K^0 \Sigma^{*+}(X)\to \pi^+\pi^-\pi^+n(X)$, it is necessary to determine which
$\pi^+$ comes from the $K^0$.  It is possible to check both final state $\pi^+$'s with the detected $\pi^-$ to study the kaon candidates in each case by using the invariant mass.  

\begin{figure}[h]
\epsfig{file=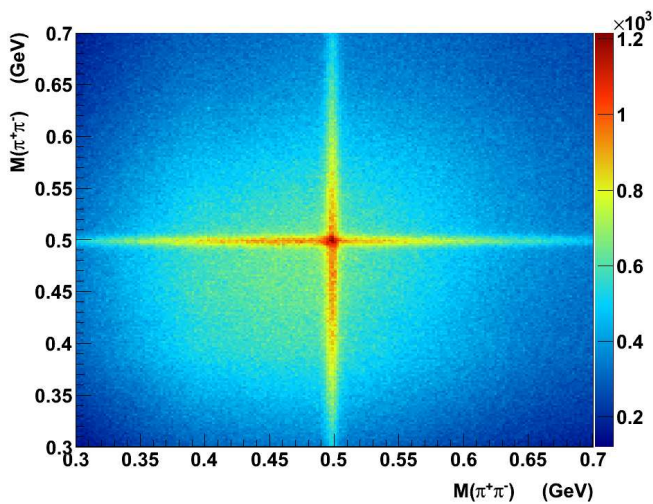,width=\columnwidth}
\caption{Invariant mass of the $\pi^+$-$\pi^-$ combination for the two different $\pi^+$ detected, prior to any $\pi^+$ organization.}
\label{k0}
\end{figure}

The invariant mass was selected for each $\pi^+\pi^{-}$ pair, as shown in Fig. \ref{k0}.  Whichever $\pi^+$ lead to the invariant
mass that was closest to the mass of the $K^0$ was associated with the $K^0$ identification.
Afterwards a cut at $\pm0.01$ GeV about the $K^0$ mass was used to clean up the selection.  For cases where both $\pi^+$ combinations with the $\pi^-$ fell within the $K^0$ mass limit
the wrong $\pi^+$ could be selected.  Monte Carlo was used to check the frequency of this ambiguity and was
found to be $\sim2\%$ for the $\gamma p \to K^{0} \Sigma^{*+}\to \pi^{+}\pi^{-}\pi^{+}n \pi^0$ and
$\gamma p \to K^{0} \Sigma^{*+}\to \pi^{+}\pi^{-}\pi^{+}n \gamma$ channels.  With additional kinematic
constraints these ambiguous events were ultimately rejected.
\subsection{Neutron identification}
Neutral particles are detected in CLAS as clusters in the electromagnetic calorimeter (EC) \cite{ec}
not associated with any reconstructed charged track from the drift chambers.  The momentum reconstruction
depends on the path length and TOF of the neutron.  The directional components of the neutral track were
found by using the vertex and the cluster position on the EC for that hit.  In this experiment the information
about the neutral vertex was limited to the information that could be extracted from the other charged
particle vertices in the decay chain.  

The EC has six triangular sectors made of alternating
layers of lead and scintillator. Scintillator layers compose
of about 10-cm-wide scintillator strips, where strips in
every consecutive layer run parallel to one of the three
sides of the triangle. The EC has 13 layers of scintillator
strips for each of the three directions making 39 total layers.
In each direction the EC is subdivided into an inner stack
of 5 layers and an outer stack of 8 layers.

The EC reconstruction software forms a
cluster by first identifying a collection of strips in each of the three views.  The software
requires a set of threshold conditions to be met and that the strips to be contiguous.  The groups of strips that pass these conditions define a $peak$ and are organized with respect to the sum of the strip energies.
The peak centroid and RMS in each of the three views
is obtained and clusters are identified as intersection of centroids
of peaks within their RMS.
If a given peak contributes to multiple hits, then the energy in each hit due to that peak is calculated as being proportional to the relative sizes of the multiple hits as measured in the other views. For example, if there are multiple hits which have the same U peak, the energy in V and W is added for each of the hits, and the ratio of these summed energies determines the weight of the U peak's energy of the multiple hits. If the software thresholds for the scintillator strip, peak and weighted hit energy are met then the cluster position and time are recorded.
The events EC time (or EC time-of-flight) is defined as the time between the event start time
and the time of the EC cluster.

During analysis the strip information was used to determine whether
the centroid was reconstructed using only the outer stack of the EC or both the inner stack
and outer stack.  The centroid could be located in any one of the layers of each stack,
however, the cluster reconstruction position did not contain that information, so the hit was assumed to 
be on the upstream face (closer to the target) of whichever stack the hit was contained in.  With the
assumed reaction vertex and the EC cluster position, the directional components in $\theta$ and $\phi$
 were found, as well as the path length of the neutron.  Using the EC time-of-flight the momentum was 
 calculated.  The neutrons were differentiated from photons using a $\beta<0.9$ cut.

Neutron detection is essential for the reaction 
of interest.  The neutron momentum was used in combination with the 
$\pi^+$ not associated with the $K^0$, to study the kinematics of the $\Sigma^+$.  Having clean constraints on the 
$K^0$ and $\Sigma^+$ is important when considering the event topology $\gamma p \to K^0 \Sigma^{*+} \to K^0 \Sigma^{+}(X)$.

A thorough study of the accuracy of the EC for neutron reconstruction
in all kinematic ranges has not been achieved previously at CLAS.  Obtaining the resolution in all measured variables for neutron reconstruction was an essential part of the present analysis.
Correlations between each measured variable in the EC had also not been  previously studied.  The
EC covariance matrix of the neutron can give a lot of information about the quality of the kinematic variables
in each region of the EC.  These values can then be used to weight the neutron measurements appropriately in kinematic constraints that depend on maximum likelihood methods \cite{keller3}.  

There are resolution differentials in all measured variables that are related to the acceptance of the EC.  Hits from the center of each triangular sector have better measurements over those on the edges due to
shower leakage.  The inner and outer stacks can act as separate detectors in the sense that if a hit
is seen in the outer but not the inner stack, then the inner stack plays no role in the reconstruction of that hit.
It is far less common for an event to pass through the inner stack with no effect and to register a hit in the outer stack, but for these events, the outer EC stack was used independently with its own unique resolution parameters for each measured variable.
All possible combinations of the measured neutron dependence on $\theta$, $\phi$, and $p$ were studied to develop a complete understanding of the neutron variance and covariance in the EC \cite{keller2}.
  
\subsubsection{Neutron detection test}

The test reaction $\gamma p \to \pi^{+}\pi^{-}\pi^{+}n$ was isolated in the $g11a$ data set by selecting
a $\pi^-$ and two $\pi^+$, and kinematically fitting to a missing neutron hypothesis and then taking
a $10\%$ confidence level cut.  Only the detected neutrons found in a direction less than $3^\circ$ from the
kinematically fit three pion missing momentum were used to ensure the correct neutron.  This channel was selected because the final decay products are identical
to the reaction of interest $\gamma p \to K^0 \Sigma^{*+} \to K^0 \Sigma^{+}(X)$.  In addition, the
momentum range of the detected particles is the same.  Kinematic constraints were imposed to remove possible $\pi^+$ $\pi^-$ combinations with invariant mass equal to the $K^0$, so that only 
the $\gamma p \to \pi^{+}\pi^{-}\pi^{+}n$ events survived.  The simplification made by working with the
test channel is that in the $\gamma p \to \pi^{+}\pi^{-}\pi^{+}n$ reaction, there is only one
interaction vertex. This implies that the neutron comes from the primary interaction vertex,
which can be well determined using the charged pions.

To study the measured neutron variable residuals, we required
each event to have one detected neutron and then compared the measured variable with the kinematically fit missing variable in each case.  Assuming a high quality missing neutron four vector, this procedure was
used to find the change in resolution with respect to all measured variables over the EC face \cite{keller2}. 
Only the events that registered an actual hit in the EC were used to study the resolution.  No
EC fiducial cuts were applied during the covariance investigation so that the entire EC face could be studied
and compared to Monte Carlo.  During analysis, only the minimal fiducial cuts were applied
of $8^\circ<\theta<40^\circ$ on the neutron polar angle to maximize the statistics.
 
For the test channel the neutron vertex was found from a multi-track-vertex fitting procedure
to give an accurate vertex (at less than 4\% uncertainty in position for the topology of interest) for multiple final state particles all coming from the same vertex \cite{mcnabb}.  Because the neutron came from the primary interaction vertex in this study, its vertex was accurately known.  However, for events in
which the neutron comes from secondary vertices, its vertex is not as easily obtained.
Because the neutron vertex information can affect its reconstructed four momentum,
these differences can be important when studying resolutions.

Once the EC neutron covariance matrix for $\gamma p \to \pi^{+}\pi^{-}\pi^{+}n$ was well understood, the Monte Carlo resolution was matched to the data using the same test channel \cite{keller2}.  The Monte Carlo was then used to study the $\gamma p \to K^0 \Sigma^{*+} \to K^0 \Sigma^{+}\pi^0$ channel and to find the neutron covariance matrix specific for this topology.  In this way the $K^0$ interaction point with the beam line could be used as the starting point of the neutron path in the neutron reconstruction process for any of the decays, so no bias was introduced by assuming a $\Sigma^{+}$.
This step removed the explicit dependence on the neutron vertex.  The Monte Carlo covariance matrix for $\gamma p \to K^0 \Sigma^{*+} \to K^0 \Sigma^{+}\pi^0$ was then used to tune the data neutron covariance matrix specific to the topology.  The change in the momentum resolution from this tuning process was smaller than $5\%$.

In order to obtain a consistent covariance matrix for the neutron, discrimination was made for each neutron between the inner and outer EC stacks
in order to calculate the correct path length.  In addition, timing and momentum corrections were applied as described below. 

\subsubsection{Neutron path}
As previously stated, the distance that the neutron travels in CLAS was used with the EC time-of-flight to determine the momentum of the neutron.
The distance is dependent on the EC stack and the position of the cluster reconstruction.  The inner stack cluster reconstruction was always used
unless there was only a hit in the outer
stack.  A determination of whether there was a hit only in the outer stack, the inner stack, or both, was
made by checking which EC scintillators were associated with an event.
The probability to find a hit in the outer stack alone was less than 15\%.
For all other neutral hits either the inner stack or both were associated with the hit.
If there was a hit only in the outer EC stack, the first layer (layer closest to the target) of the outer stack gave the plane of the EC cluster coordinates.  If there was a hit in the inner stack or both, the first layer of the inner stack was used as the plane of the EC cluster coordinates. 

\subsubsection{Neutron time}
The time-of-flight for the neutron came from the difference between the event start time and the EC
cluster time.  The path length used to reconstruct the neutron momentum, which assumes a hit on the EC face of either the inner or outer stack, was inaccurate by the distance the
neutron traveled past the EC face into the detector.  A correction was used
to compensate for the average additional distance the neutron travels into the EC.
In addition the outer stack is farther from the target and for the same event would have
a slightly different time response than that of the inner stack.

A correction was implemented directly in the neutron time-of-flight to correct the neutron momentum. This was
done by using the calculated neutron time-of-flight and comparing it to the expected time
$T_{expected}=L/c\beta_{miss}$.  Here,
$L$ is the path length of the neutron, and $\beta_{miss}$ is the $\beta$ calculated using the missing
momentum and energy of the neutron from the $\gamma p \to \pi^+\pi^-\pi^+(n)$ events
that passed a 10\% confidence level cut from a kinematic fit under a missing neutron hypothesis.  By using $T_{expected}-T_{EC}$ for each stack,
a separate correction was found for each case, such that $T_{expected}-(T_{EC}+T_{correction})\sim0$.  By finding
separate timing corrections for the inner and outer stacks, the farther distance of the outer stack was compensated for.
The separate study of timing corrections for the inner and outer stacks was carried out using the Monte Carlo simulations.

The timing correction used is the same in all directions.  However, in order to obtain
accurate covariance information an additional momentum correction was required which is sensitive to the geometry of the EC and neutrons trajectory.  It was only after all corrections that the residual means of all measured variables were centered around zero to accurately reflect the neutron resolutions. 
   
\subsubsection{Neutron momentum correction}
A neutron momentum correction was implemented by studying the trend found in
momentum and position resolutions over various kinematic ranges.  This was done by studying the residuals
$\Delta p$, $\Delta \theta$, and $\Delta \phi$ over each variable $p$,  $\theta$,
and $\phi$.  The residual of momentum, $\Delta p$, is defined as the difference between the missing neutron momentum
and the reconstructed neutron momentum.  Likewise for the directional components $\Delta
\theta$ and $\Delta \phi$.  The missing neutron four-vector was found by kinematically fitting the charged decay products in the missing mass of the neutron and taking a 10\% confidence level cut. In this fit there were three unknowns from the components of the missing momentum vector and four constraints from the conservation of energy and momentum to make a 1-C fit \cite{keller3}.
The trend of each of the residuals
should be distributed around zero, if it is not, the distribution will display a trend
that can be used to correct the measured variable.  Once the neutron momentum 
magnitude and directional resolutions are evenly distributed around zero, the 
missing and detected four-vectors are comparable.  This implies that for the 
majority of events, the detected neutron momentum vector was the same within the experimental
resolution as the high quality kinematically fit missing neutron momentum vector.

\begin{figure}
\epsfig{file=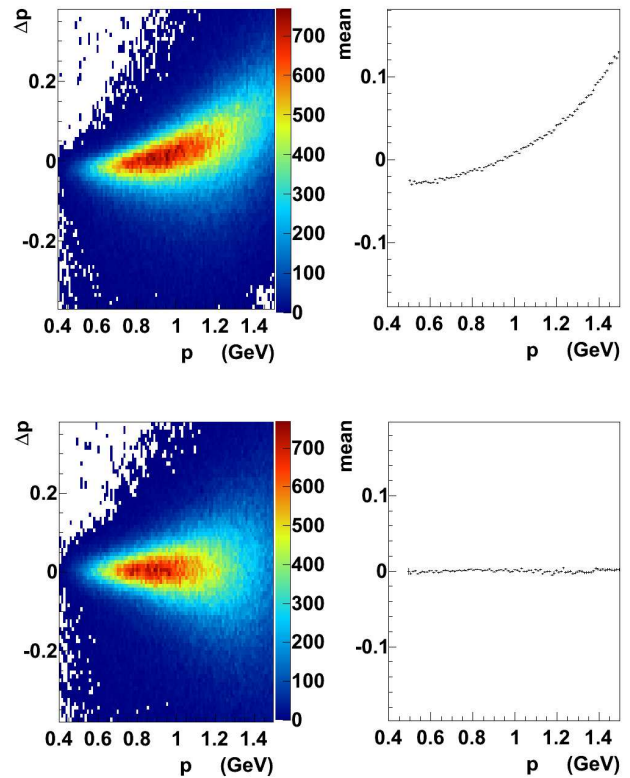,width=\columnwidth}
\caption{Top left: the neutron $\Delta p$ distribution before correction,
top right: the Gaussian mean from the binned fits, bottom left: the corrected
distribution, and bottom right: the final corrected Gaussian mean from the
binned fits.}
\label{mom_cor}
\end{figure}

Similar corrections were implemented to $\Delta p$ with respect to $\phi$ such that
$p'=p+f(\phi)$, $\Delta \phi$ with respect to $\phi$ such that $\phi'=\phi+f(\phi)$, and
$\theta$ with respect to $\theta$ such that $\theta'=\theta+f(\theta)$. 
The Monte Carlo required separate corrections in the same
variables that were determined using the same procedures as for the
data.
\subsubsection{Neutron covariance matrix}
The neutron covariance matrix was determined after the corrected neutron path was used with
the timing correction implemented for the corresponding EC stack and momentum correction.
This covariance matrix was required in order to kinematically fit the neutron with the other
detected particles.  The variables used to represent the neutron vector components were $\theta$, $\phi$, and $p$, leading to a covariance matrix of
\begin{eqnarray}
{\rm\bf V}^{n}_{i} =
\left( \begin{matrix}
    V^{pp}_{i}       &      V^{p\theta}_{i}         &      V^{p\phi}_{i}     \cr
    V^{p\theta}_{i}  &      V^{\theta\theta}_{i}    &      V^{\phi\theta}_{i} \cr
    V^{p\phi}_{i}    &      V^{\phi\theta}_{i}      &      V^{\phi\phi}_{i}   \cr
\end{matrix}\right). \nonumber
\end{eqnarray}
The variance and correlations of each variable were obtained by studying the differences between the kinematic variables of the detected neutron from the kinematically fit missing neutron.  The
residuals in each case were sliced and binned to fit with a Gaussian to
find the functional dependence in each variable.  Once the functional dependence on each variable 
was found for each $\phi$ and $\theta$ bin, an empirical smearing technique was used to get the Monte Carlo to closely match the same functional dependence seen in the data.  Similar steps were taken for the directional components \cite{keller2}.  

\section{Analysis Procedure}
\label{pro}
In the analysis, progressive steps were taken to remove as much identifiable background as possible while preserving the counts from the channels $\gamma p \to K^{0} \Sigma^{*+} \to K^0 \Sigma^{+}(\pi^0)$ and 
$\gamma p \to K^{0} \Sigma^{*+} \to K^0 \Sigma^{+}(\gamma)$.  The radiative signal
was buried by the $\pi^{0}\to \gamma \gamma$ decay and required advanced fitting techniques to 
resolve the signal.  The fitting procedure developed here required that all other backgrounds be removed
or extensively minimized to ensure high quality separation between the radiative and strong decays of the $\Sigma^{*+}$.

\begin{figure}
\epsfig{file=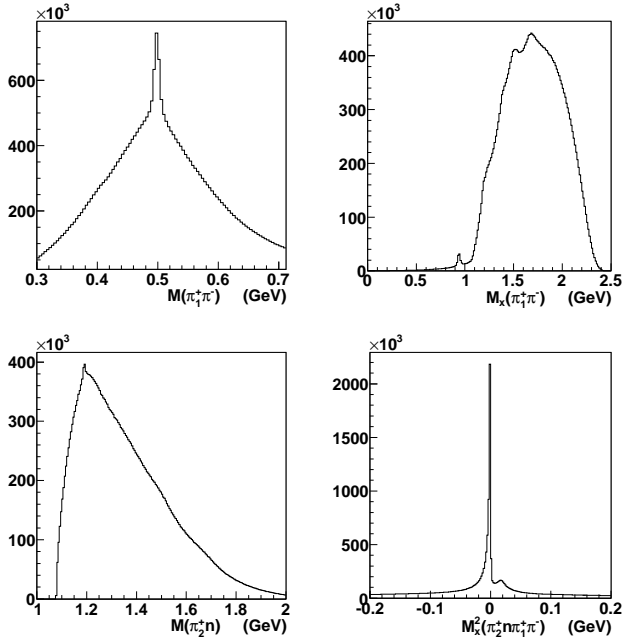,width=\columnwidth}
\caption{The invariant mass of the $\pi_1^+$-$\pi^-$ (upper left), the missing mass off the $\pi_1^+$-$\pi^-$
(upper right), the n-$\pi_2^+$ invariant mass (lower left), and the missing mass squared of all the detected
 particles (lower right).  All distributions are before any kinematic constraints.}
\label{fig1-spec_1}
\end{figure}

\begin{figure}
\epsfig{file=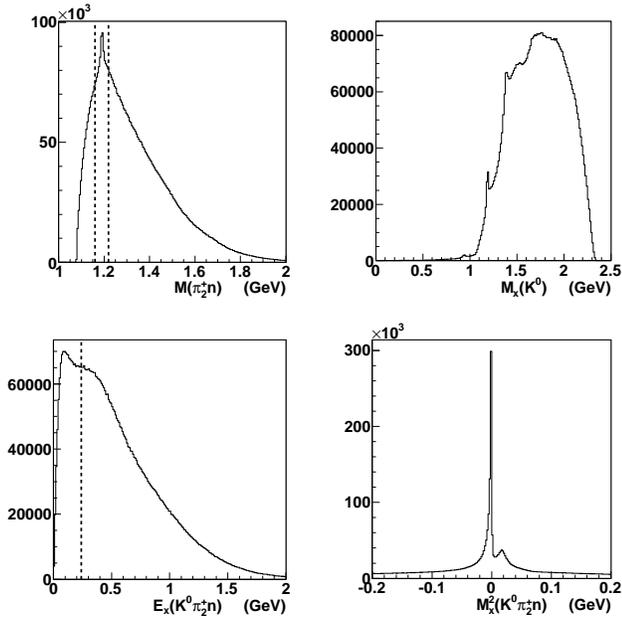,width=\columnwidth}
\caption{The invariant mass of the n-$\pi_2^+$ with the dotted lines showing the $|M(\pi_2^{+}n)-M_{\Sigma^{+}}|<0.01$ GeV cut (upper left), the missing mass off the $K^0$ (upper right),
the total missing energy with the dotted line showing the cut used (lower left), and the missing mass squared of all the detected particles (lower right) after
the $\pm0.01$ GeV cut on the $K^{0}$ peak.}
\label{fig2-spec_2}
\end{figure}

The goal was then to achieve clean hadron identification
before using the fitting procedure for the competing $\pi^{0}$ and radiative signals.
For the sake of notation, let $\pi_1^{+}$ indicate the $\pi^+$ used in the $K^{0}$ invariant mass selection, such that $\pi_1^{+}$ is the $\pi^+$ that forms the closest known $K^{0}$ mass
when combined with the $\pi^-$.
Naturally $\pi_2^{+}$ is the other detected $\pi^+$.  Fig. \ref{fig1-spec_1} shows
the invariant mass of the $\pi_1^+$-$\pi^-$ (upper left), missing mass off the $\pi_1^+$-$\pi^-$ (upper right),
the n-$\pi_2^+$ invariant mass (lower left), and the missing mass squared of all the detected
particles (lower right).  The distributions in Fig. \ref{fig1-spec_1} are before any kinematic
constraints and after the $\pi^+$ assignments are made.
The $K^{0}$ was cut about $\pm0.01$ GeV of the known $K^{0}$ mass to reduce the $\pi^+$-$\pi^-$ background.  Fig. \ref{fig2-spec_2} shows
the invariant mass of the n-$\pi_2^+$ (dashed lines show the cut that was implemented) (upper left), missing mass off the $K^0$ (upper right),
the missing energy off all detected particles  (dashed lines show the cut that was implemented) (lower left), and the missing mass squared of all
the detected particles (lower right), after the $K^0$ cut.
The $\Sigma^{+}$ peak is clearly visible as seen in the (upper left) plot.  The
clear visible peak in the missing mass squared at the $\pi^0$ mass is also an indication that the neutron measurement is effective.

\begin{figure}
\epsfig{file=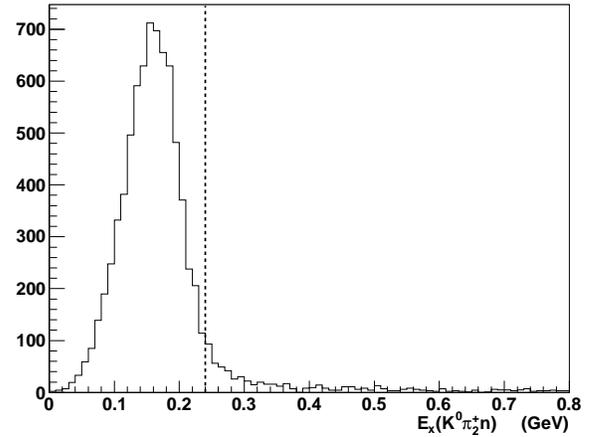,width=\columnwidth}
\caption{The Monte Carlo missing energy distribution for the $\gamma p \to K^{0} \Sigma^{*+} \to K^0 \Sigma^{+}\gamma$ reaction with the dotted line indicating the 0.24 GeV cut.}
\label{Ei_gam}
\end{figure}

\begin{figure}
\epsfig{file=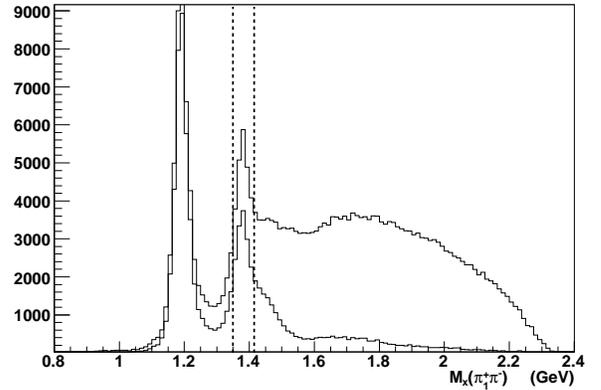,width=\columnwidth}
\caption{The missing mass off the $\pi_1^+$-$\pi^-$ with three progressive cuts applied
to isolate the $\Sigma^{*+}$ events.  First $|M(\pi_2^{+}n)-M_{\Sigma^{+}}|<0.01$ GeV, next shown are the events
left over after the $E_{x}<0.24$ GeV cut, and finally the dotted lines show the $\pm$0.03 GeV cut around the
mass of the $\Sigma^{*+}$.}
\label{fig3-spec}
\end{figure}

A Monte Carlo study on the phase space of the $\gamma p \to K^{0} \Sigma^{*+} \to K^0 \Sigma^{+}(X)$ reaction
indicated that most events from the missing energy boosted in the $\Sigma^{*+}$ frame $E_x$ should be
in the range of 0-0.25 GeV.  A cut at 0.24 GeV was chosen to clean up the $\Sigma^{*+}$
candidates.  This cut preserved $\sim80\%$ of the radiative and $\pi^0$ signals, while
substantially reducing the background under the $\Sigma^{*+}$.  Fig. \ref{Ei_gam} shows
an example of the Monte Carlo missing energy distribution for the $\gamma p \to K^{0} \Sigma^{*+} \to K^0 \Sigma^{+}\gamma$ reaction with the dotted line indicating the 0.24 GeV cut. 
Fig. \ref{fig3-spec} shows the results on the missing mass off the $K^0$.
A cut on the invariant mass of the n-$\pi_2^+$ combination along with the missing energy
cut cleans up the excited-state hyperon spectrum, making the $\Sigma^{*+}$ quite prominent.  Finally
a $\pm 0.03$ GeV cut was applied to the missing mass off the $K^0$ around the known mass of the
$\Sigma^{*+}$. 

\begin{figure}
\epsfig{file=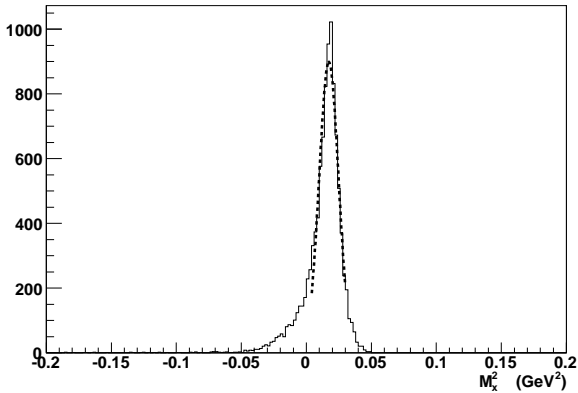,width=\columnwidth}
\caption{The missing mass squared of all detected particles after all analysis cuts.  A Gaussian fit
gives a mean of 0.018$\pm$0.0002 GeV.}
\label{fig4-spec_3}
\end{figure}

\begin{figure}
\epsfig{file=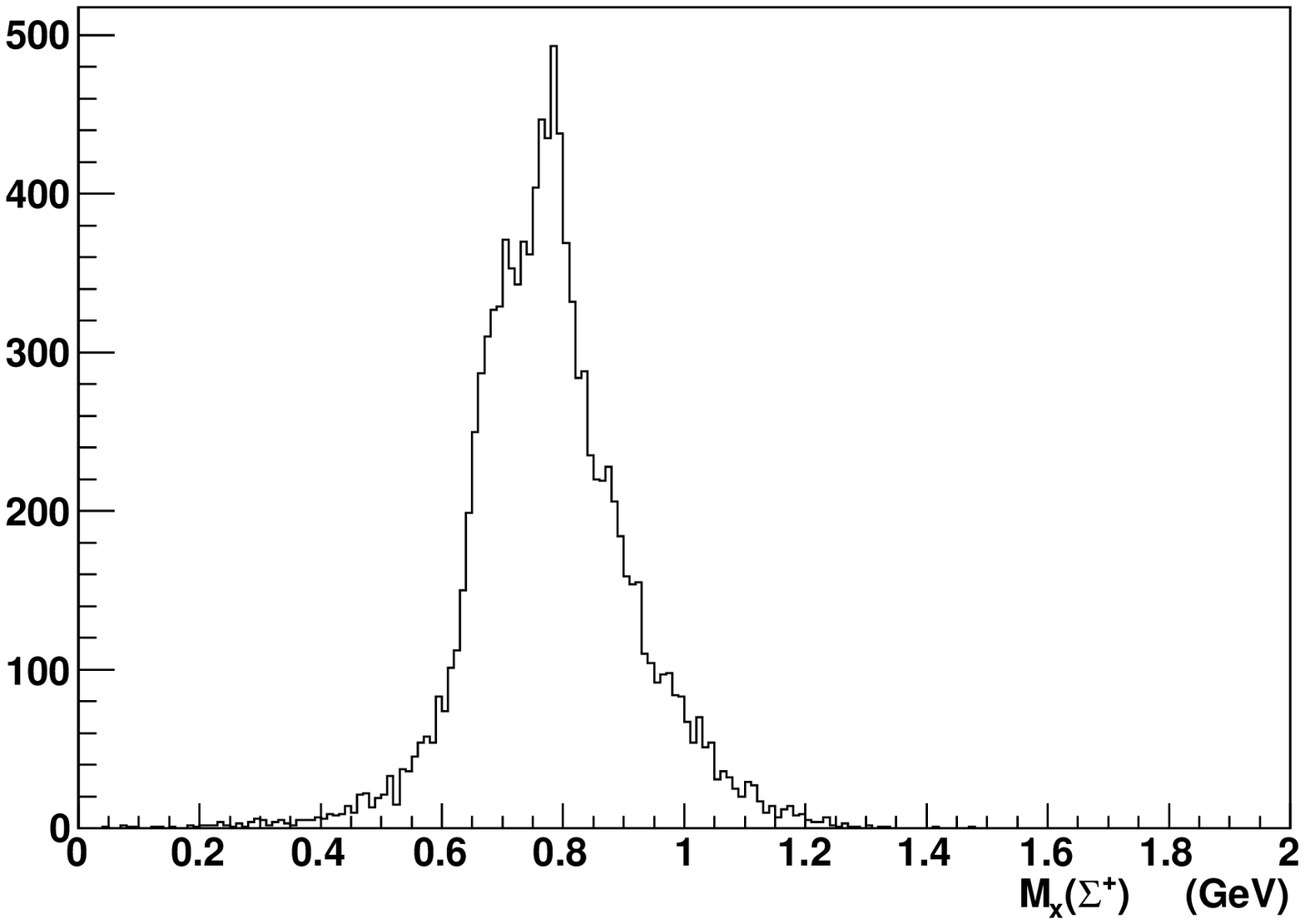,width=\columnwidth}
\caption{The missing mass of the $\Sigma^{+}$.}
\label{fig5-spec_4}
\end{figure}

Fig. \ref{fig4-spec_3} shows the missing mass squared of all detected particles after all of the mentioned cuts.
A clear $\pi^0$ peak is present with some smaller but unknown amount of radiative signal at zero missing mass.
Fig. \ref{fig5-spec_4} shows the the missing mass off the n-$\pi_2^+$ combination.  The missing mass off the $\Sigma^{+}$ will be used in the background analysis.


\subsection{Simulations}
\label{sims}

A Monte Carlo simulation of the CLAS detector was performed using 
GEANT \cite{geant}, set up for the $g11a$ run conditions.
The experimental photon energy distribution for an incident electron beam of 4.0186 GeV was used
to determine the energies of the incident photons in the simulation.
Events were generated for the radiative channel 
($\Sigma^+(1385) \rightarrow \Sigma^+\gamma$),
the normalization reaction 
($\Sigma^+(1385)\rightarrow\Sigma^+\pi^0$), 
and several background reactions. 

A phase space event generator was used with a variable $t$-dependence such that a channel with a $K^0$ was generated uniformly in the center-of-mass
frame in $\phi$ with a $t$-dependent distribution in $\theta_{cm}$ according to $P(t) \propto e^{bt}$ with $b$=2.0~GeV$^{-2}$.
Gaussian distributions in $x$ and $y$ with $\sigma=0.5$ cm were used to approximate the beam width
at the target.
Events were generated uniformly along the length of the target.
These generated events were fed into a simulation of the CLAS detector.  

For each contributing channel the differential cross section was found using data to weight the strength and angular distribution in the Monte Carlo generator.
A very careful empirical smearing procedure was used to match the Monte Carlo and data resolutions.  This procedure is discussed in Refs. \cite{keller3} and \cite{keller2}.  Ultimately the missing mass squared from Monte Carlo,
$M_x^2(K^0\Sigma^+)$, gave very good agreement with the shape of the
experimental data as shown in Fig. \ref{fig4-spec_3}.

This analysis relies on an understanding of the contributing leakage of background channels
into the $\pi^0$ and radiative signal peaks.  For example, $\pi^0$
leakage from a background channel such as $\gamma p \to \omega \Delta^{+} \to \pi^+\pi^-\pi^0 n \pi^+$
will lead to over-counting of $\Sigma^{*+} \to \Sigma^+ \pi^0$ events.  The Monte Carlo of various possible contributions was used to study the possible background 
leakage at various stages of the analysis.  The acceptances of each possible background were used to 
study the possible effect on the final reported ratio.

At this stage of the analysis the most likely background reaction is $\gamma p \to \omega \Delta^{+}$, 
followed by $\Delta^+ \to n \pi^+$ decay.  The $\omega$ decays 
primarily to $\pi^+\pi^-\pi^0$ followed by $\pi^0 \to 2\gamma$. 
The full reaction $ \gamma p \to \omega \Delta^{+}$ then has the same 
final state as $K^{0}\Sigma^{*+}\to\Sigma^{+}\pi^{0}$ and must be
carefully considered.  This is also true for reactions like $\gamma p \to \omega N^{*}$.
The $N(1440)$ has a relatively large decay width
at 250-450 MeV \cite{pdg}, which implies possible leakage into any cut around the $\Sigma^{*+}$.  This is 
the reason for the extra steps to isolate the $\Sigma^{*+}$ as shown in Fig. \ref{fig3-spec}.
Due to the constraints on the $\pi^+$ and $n$, along with the series of cuts shown
in Fig. \ref{fig3-spec},
the leakage from the $N^{*}$ decay was negligible.  However, because the $\Delta^{+}$ is very close in mass
to the $\Sigma^+$ and the $\pi^+\pi^-$ from the $\omega$ decay has a similar phase space to the $K^0$,
there was some $\gamma p \to \omega\Delta^{+}$ leakage that needed to be accounted for.

The contribution of the $K^{*0}$ background was also studied.  The constraints on the
reconstructed $K^{0}$, combined with the missing mass constraint off the $K^{0}$
to be the mass of the $\Sigma^{*+}$, should minimize any $K^{*0}$ contribution.
However, because the reaction $\gamma p \to K^{*0} \Sigma^{+}$ has the same
possible final states that are being analyzed, it was carefully considered.  The Monte Carlo
investigation indicated that there were contributions that needed to be accounted for.

Also investigated with Monte Carlo was the reaction
$ \gamma p \to m^* N^*$, where $m^*$ is any meson that can decay to $\rho\pi^0$, and 
the $N^* \to N\pi^+$ provides the detected pion. Similarly, $ \gamma p \to \rho N^*$,
where the $N^*$ decay to $n\pi^+\pi^0$ was a possible contaminant.  In addition to 
the kinematic constraints previously mentioned, these backgrounds 
cannot contribute for low $W$ ($W<1.6$ GeV).  For testing purposes of these types
of reactions, the channel $\rho N(1520)$ was considered.  The $\rho(770)$
has a width of $\Gamma = 150.3$ MeV and decays almost 100\% to $\pi \pi$, so it
was possible to leak under the $K^0$ invariant mass cut.  Ultimately all contributions
of the channel type $ \gamma p \to m^* N^*$ were found to be negligible (zero acceptance).

Based on the possible final state decay products, the reactions $\gamma p \to \eta n\pi^{+}$,
$\gamma p \to K^0 \Sigma^0 \pi^+$, and $\gamma p \to K^0 \Sigma^+$ were also considered.
These backgrounds also have negligible acceptance as determined from high-statistics Monte Carlo studies using the same event selection as for the data, 
and hence were dismissed.

\subsubsection{Minimization of the $\omega$ and $K^{*0}$ backgrounds}
\label{bgRemoval}
\begin{figure}
\epsfig{file=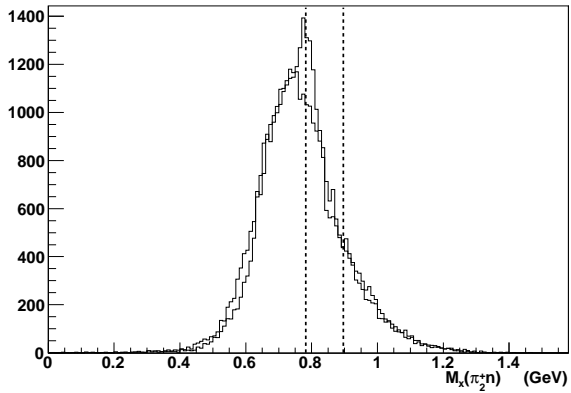,width=\columnwidth}
\caption{The Monte Carlo and data distributions for the missing mass off the $\pi_2^{+}n$ combination with the data showing a $\omega$ peak.  The Monte Carlo shown is for the reaction $\gamma p \to K^{0}\Sigma^{*+} \to \pi^+\pi^-\pi^+n\pi^0$, indicating the expected missing mass off the $\pi_2^{+}n$ combination if no background was present.  The dashed lines indicate the known masses of the $\omega$ and $K^{*0}$.}
\label{MCtoData_a}
\end{figure}

As indicated in the previous section, the $\gamma p \to \omega\Delta^+$ and $\gamma p \to K^{*0}\Sigma^+$ channels are
the most likely background contributors.  The branching ratio of $\omega \to \pi^+ \pi^- \pi^0$ is
$89.2\pm0.7\%$ \cite{pdg}, implying a high probability of overlap with the normalization channel $\gamma p \to K^0 \Sigma^{*+} \to \pi^+ \pi^- \pi^+ n (\pi^0)$.  The $\gamma p \to K^{*0}\Sigma^+$ channel was a concern for the same reason.
To get an indication of how much these channels were present in the data, the missing mass off the $\pi_2^+$-$n$ combination
was used.  For the $\gamma p \to \omega\Delta^+$ ($\gamma p \to K^{*0}\Sigma^+$) channel the missing mass off $\pi_2^+$-$n$ should show a $\omega$ ($K^{*0}$) peak.
The missing mass spectrum off the $\pi_2^+$-$n$ combination from the Monte Carlo of the $\gamma p \to K^0 \Sigma^{*+} \to \pi^+ \pi^- \pi^+ n (\pi^0)$ channel was compared to the same distribution from data.  To isolate the $\pi^0$ channel in the data, a kinematic fit to the missing $\pi^0$ with a $10\%$ confidence level cut was applied to leave only the final state $\pi^+ \pi^- \pi^+ n \pi^0$ in the data.  A direct comparison between the data and Monte Carlo 
of the missing mass spectrum off the $\pi_2^+$-$n$ combination then deviated where background was present (normalizing the Monte Carlo to the data). 
It is clear from the comparison shown in Fig. \ref{MCtoData_a} that there is a non-negligible amount of $\omega$ events.  The number of events with a $K^{*0}$ present is too small to be visible.
\begin{figure}
\epsfig{file=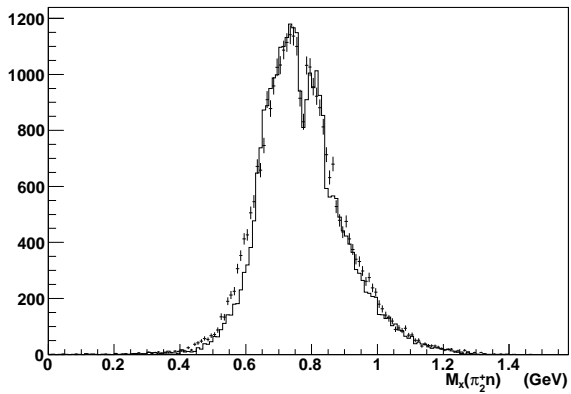,width=\columnwidth}
\caption{The Monte Carlo (lines) and data (points) distributions for the missing mass off the
$\pi_2^{+}n$ combination (as shown in Fig. \ref{MCtoData_a}) after the $P(\chi^2)<1\%$ cut showing that the distributions now match
in the $\omega$ mass region.
}
\label{MCtoData_b}
\end{figure}  

To remove the majority of the $\omega$ events a kinematic fit was performed with a missing $\pi^0$ hypothesis, while constraining the $\pi_1^+,\pi^-$ and $(\pi^0)$ to
have the $\omega$ mass, resulting in a 2-C fit.  High confidence level candidates were then rejected as
part of the identifiable $\omega$ background.  Various confidence level cuts were tested until the data 
matched the Monte Carlo in the mass range of the $\omega$ (within the statistical error bars of the data).  
Ultimately a confidence level cut of $P(\chi^2)<1\%$ was used, resulting in the comparison seen in 
Fig. \ref{MCtoData_b}.  The same cut was used to reduce the possible $\gamma p \to \omega\Delta^+\to\pi^+ \pi^- \pi^+ n (\gamma)$
background by imposing the constraint on the $\pi_1^+,\pi^-$ and $(\gamma)$ to be $\omega$.  In this case it was not
possible to use the Monte Carlo and data to check in the same way, and so a $P(\chi^2)<1\%$ cut was used.  The same cut was also used
under a $K^{*0}$ hypothesis to reduce the acceptance of the $\gamma p \to K^{*0}\Sigma^+\to \pi^+ \pi^- \pi^+ n (\pi^0)$ channel.  Similarly for 
the $\gamma p \to K^{*0}\Sigma^+\to \pi^+ \pi^- \pi^+ n (\gamma)$ with the hypothesis of the $\pi_1^+,\pi^-$ and $(\gamma)$ to be $K^{*0}$.

Even with the above cuts in place, a small amount of the $\omega$ and $K^{*0}$ background still slipped through.  An estimate of
this leakage was found and then subtracted out of the final result, as discussed in the following
sections.

\subsubsection{Cross sections}
\label{cross}
To tune the $\Sigma^{*+}$ Monte Carlo, the differential cross sections for the reactions $\gamma p \to
K^0\Sigma^{*+} \to \pi^+ \pi^- n \pi^+\pi^{0}$, $\gamma p \to K^{*0}\Sigma^{+} \to \pi^+ \pi^- \pi^{0} n
\pi^+$, and $\gamma p \to \omega\Delta^+ \to \pi^+ \pi^- \pi^{0} n \pi^+$ were obtained.  The shapes of the
differential cross sections were then used to adjust the event generators. 
In each case a $1/E_{\gamma}$ photon energy distribution was used in the generator. 

A normalization procedure shown to accurately reproduce a number of well-measured channels was used for each cross section \cite{mccracken}.
The following is a discussion of the procedure used to extract the $\gamma p \to K^0\Sigma^{*+}$ cross section.  A similar procedure was followed for the two background channels.

\begin{figure}
\epsfig{file=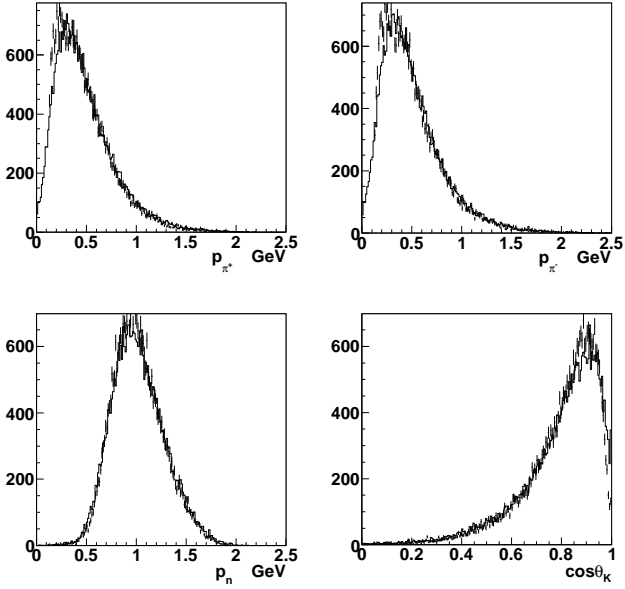,width=\columnwidth}
\caption{The momentum distributions for the $\pi^+$, $\pi^-$, and neutron are shown in the first 3 panels, 
with the cosine of the kaon lab frame angular distribution in the last panel, 
for the reaction $\gamma p \to K^0\Sigma^{*+}\to\pi^- \pi^+ n \pi^+ \pi^0 $.
The data and Monte Carlo are shown as the histogram and points, respectively, 
which closely overlap.
}
\label{momCom}
\end{figure}

With all of the aforementioned constraints the $\gamma p \to K^0\Sigma^{*+} \to K^0\Sigma^{+}\pi^0$ reaction was 
easily isolated with a kinematic fit to the missing $\pi^0$.  A $10\%$ confidence level cut was applied
to ensure channel purity.  The yield was determined by the ratio of the raw $\Sigma^+ \pi^{0}$ events to the number of incident photons in each $E_{\gamma}$
bin, so as to normalize with the bremsstrahlung spectrum.  Corrections were made for each bin with the newly obtained acceptances.  The angular dependence in
the generator was initially flat with a zero $t$-slope dependence.  After the differential cross section was obtained, the distributions were used to adjust
the $\Sigma^{*+}$ generator.  Each corresponding angle and energy bin was filled according to the distributions seen in the data results.  Each angle bin was
divided into $E_{\gamma}$ bins and represented accordingly in the new event weighting scheme of the generator.  The adjusted generator was then used to
produce new Monte Carlo and obtain more accurate acceptances.  This process was then iterated until no
change was seen to the differential cross sections within the statistical uncertainties.
After these modifications were made, the resulting Monte Carlo was compared with the data, using the momentum distributions for the $\pi^-$, $\pi^+$, and neutron tracks, as well as the $K^0$ lab frame angle distribution, and found to closely match the data within the statistical uncertainties, see Fig. \ref{momCom}.

The same corrections were applied to the $\gamma p \to K^0\Sigma^{*+} \to K^0\Sigma^{+}\gamma$ cross sections.
The nature of the corrections to the Monte Carlo were specific to the $\gamma p \to K^0\Sigma^{*+}$ cross section
and so the corrections could be applied without discrimination between the electromagnetic and strong decays of the baryon.

To calculate the acceptance of the signal and background reactions, an extraction method used to resolve the radiative and $\pi^0$ channels was required and will be discussed next.


\subsection{Fitting technique}
\label{ExtMeth}
The two-step kinematic fitting procedure developed in Ref. \cite{kellpaper} was employed to resolve the radiative and strong decay signals.  Because of such similar topologies and the small relative size of the radiative signal, the kinematic fitting procedure could not be expected to cleanly separate the $\Sigma^+\gamma$ events from
the overwhelming $\Sigma^+\pi^0$ events in a single fit.  Thus we employed a two-step kinematic fitting procedure, making first a kinematic fit to a missing $\pi^{0}$ hypothesis, then checking the quality of the fit of the low confidence level candidates in a second kinematic fit to the actual radiative hypothesis.

In order to check the quality of the kinematic fit to a particular hypothesis, we studied
the $\chi^{2}$ distribution from the fitting results.  In this procedure all detected particles were kinematically fit to the
appropriate missing mass hypothesis.  An additional constraint was introduced into the kinematic fitting
enabling analysis of the more well-behaved 2-C $\chi^2$ distribution, as opposed to the 1-C distribution, to test the quality of the candidates with the hypothesis used.  The constraint required that the neutron and $\pi^{+}$ track have an invariant mass of the $\Sigma^+$ in the hypothesis. 
The detected particle tracks were kinematically fit as the final stage of analysis and filtered with the confidence level cut.
In this fit there were three unknowns ($\vec p_{x}$) and five constraint equations, four from conservation of momentum and then the additional
invariant mass condition.  This makes a 2-C kinematic fit. 
In the attempt to separate the various contributions of the $\Sigma^{*+}$ radiative decay and the decay to $\Sigma^+\pi^{0}$, the events were fit using different hypotheses for the topology:
\begin{center}
\begin{tabular}{ccc}
&$\gamma p \rightarrow \pi^+ \pi^- \pi^+ n (\pi^0)$  & 2-C\\
&$\gamma p \rightarrow \pi^+ \pi^- \pi^+ n (\gamma)$ & 2-C.\\
\end{tabular}
\end{center}
The constraint equations were
\begin{equation}
\left[ \begin{array}{c}
	(E_{\pi^{+}_2}+E_n)^2-(\vec p_{\pi^+_2}+\vec p_{n})^2-M^2_{\Sigma^{+}}\\
	E_{beam}+M_p-E_{\pi^+}-E_{\pi^-}-E_{\pi^+}-E_n-E_x\\
	\vec p_{beam} - \vec p_{\pi^+} - \vec p_{\pi^-} -\vec p_{\pi^+} -\vec p_x
	\end{array} \right]=\vec 0. 
\end{equation}
$\vec p_x$ and $E_x$ represent the missing momentum and energy of the undetected $\pi^0$ or $\gamma$. 

Then a fit function was made for a 2-C $\chi^2$ distribution following from Ref. \cite{keller3} as
\begin{equation}
f(\chi^{2})=\frac{P_{0}}{2}e^{-P_{1}\chi^{2}/2}+P_{2}.
\label{fittingf}
\end{equation}
This fit function has a flat background term, $P_{2}$.  $P_{1}$ was used to measure how close the distribution in the histogram was to the ideal theoretical $\chi^{2}$ distribution for two degrees of freedom.

Because there were two kinematic fits for both the $\pi^{0}$ and radiative channels, 
some new notation is introduced.  
The first confidence level cut used to filter out the larger $\pi^{0}$ 
signal from the radiative signal by using a kinematic fit to 
$\Sigma^+ (\pi^{0})$ and taking only the {\it low} confidence level candidates 
is denoted as $P_{\pi}^{a}(\chi^{2})$.  
The final kinematic fit used to isolate the radiative signal, using a 
$\Sigma^+(\gamma)$ hypothesis has a confidence level cut denoted as 
$P_{\gamma}^{b}(\chi^{2})$, taking only the {\it high} confidence level 
candidates.  Optimization studies have been previously done to constrain the choice of $P_\pi^a(\chi^2)$ and $P_\gamma^b(\chi^2)$ \cite{keller3}.

\subsection{Ratio calculation}
\label{MCx}
The $\pi^0$ leakage into the $\gamma$ channel was the dominant 
correction to the radiative branching ratio.  To properly calculate the
ratio, the leakage into the  $\pi^0$ region from the $\gamma$ channel was also
used.  Taking just these two channels into consideration, the number of 
$true$ counts is represented as $N(\Sigma \gamma)$ for the $\Sigma^{*+} \to \Sigma^+ \gamma$ channel and
$N(\Sigma \pi)$ for the $\Sigma^{*+} \to \Sigma^{+} \pi^{0}$ channel.  The acceptance under the $\Sigma^{*+} \to
\Sigma^{+} \gamma$ hypothesis is written as $A_\gamma(X)$, with the subscript showing the hypothesis type and the actual channel of Monte Carlo input that was used
to obtain the acceptance value indicated in the parentheses.  For the calculated acceptance of the $\Sigma^{*+} \to
\Sigma^+ \gamma$ 
channel under the $\Sigma^{*+} \to \Sigma^+ \gamma$ hypothesis, the acceptance is $A_\gamma(\Sigma \gamma)$, and for the
$\Sigma^{*+} \to \Sigma^+ \pi^{0}$ hypothesis
it is $A_\pi(\Sigma \gamma)$.  It is now possible to express the measured yields for each channel $n_{\gamma}$ and $n_{\pi}$ as
\begin{equation}
n_{\gamma}=A_{\gamma}(\Sigma\gamma)N(\Sigma\gamma)+A_{\gamma}(\Sigma \pi)N(\Sigma \pi)
\label{eq_1}
\end{equation}
\begin{equation}
n_{\pi}=A_{\pi}(\Sigma\pi)N(\Sigma\pi)+A_{\pi}(\Sigma \gamma)N(\Sigma \gamma).
\label{eq_new_2}
\end{equation}
The desired branching ratio of the radiative channel to the $\pi^{0}$ channel using the $true$ counts is then
$R=N(\Sigma \gamma)/N(\Sigma \pi)$.  Solving for $R$ to get the branching ratio expressed in terms of measured values and acceptances gives
\begin{equation}
R = \frac{n_{\gamma}A_{\pi}(\Sigma\pi)-n_{\pi}A_{\gamma}(\Sigma\pi)}{n_{\pi}A_{\gamma}(\Sigma\gamma)-n_{\gamma}A_{\pi}(\Sigma \gamma)}.
\label{eq_newR}
\end{equation}
Equation \ref{eq_newR} is based on the assumption that there are no further background contributions.
The formula for the branching ratio to take into account background from the $\omega \Delta^+$ channel, as an
example, can be expressed as
\begin{eqnarray}
R&=&{\Delta n_\gamma
 A_\pi(\Sigma\pi)
- \Delta n_\pi
 A_\gamma(\Sigma\pi) \over \Delta n_\pi A_\gamma
(\Sigma\gamma)
- \Delta n_\gamma A_\pi(\Sigma\gamma)}, \nonumber\\
\label{R}
\end{eqnarray}
where
\begin{eqnarray}
\Delta n_\pi&=& n_\pi-N_\pi(\omega\rightarrow \pi^+\pi^-\pi^0)\nonumber\\
			&-&N_\pi(\omega\rightarrow \pi^+\pi^-\gamma)
\label{aa}			
\end{eqnarray}
and
\begin{eqnarray}			
\Delta n_\gamma&=&n_\gamma-N_\gamma(\omega\rightarrow \pi^+\pi^-\gamma)\nonumber\\
			&-&N_\gamma(\omega\rightarrow \pi^+\pi^-\pi^0).
\label{bb}
\end{eqnarray}

The $n_\gamma$ ($n_\pi$) terms come directly from the yield of the kinematic fits and
represent the measured number of photon (pion) candidates.  In the notation used, 
lowercase $n$ represents the measured counts, while uppercase $N$ represents the acceptance corrected or
derived quantities.  The $N_{\gamma,\pi}$ terms are corrections needed for the leakage from 
the $\omega \Delta^{+}$ channel (an arbitrary number of background types $N_{\gamma,\pi}(X)$ can be accounted for in this manner).  The notation utilized is
such that the pion (photon) contributions are denoted 
$N_{\pi}$ ($N_{\gamma}$), so that 
$N_\gamma(X)$ denotes the relative leakage of 
the $(X)$ channel under the $\Sigma^+\gamma$ hypothesis and 
$N_\pi(X)$ denotes the relative leakage of the $X$ channel
under the $\Sigma^+\pi^0$ hypothesis.

The final acceptance for each channel was determined after
the final set of confidence level cuts was taken.  After the background acceptances were minimized, an estimate
of the background contributions was found for each relevant case.

\section{Signal Extraction}
\label{sigX}
Each Monte Carlo channel was run through the analysis with the
same cuts as used for the data.  These cuts for the extraction of the radiative and $\pi^0$ signals are listed in Table \ref{cuts_list}.
The cuts are listed in the order implemented.  The first cut, (1), was on the $K^0$ mass and restricted the $\pi^+_1,\pi^-$ sample.  The second cut, (2), on the $\Sigma^{+}$ mass was
implemented to clean up the $\pi_2^+,n$ sample prior to the more restrictive cuts, (5)-(8).  The missing energy restriction used to reduce background is number (3).
Cut (4) restricted the missing mass off the $K^0$ to be in the range of the $\Sigma^{*+}$.  Cuts (5) and (6)
list the final confidence level cuts used from the kinematic fit to isolate the missing $\pi^0$.  Cuts (7) and (8) were used to isolate the radiative decay.  The second column lists whether the cut was applied to just one channel or both.  
\begin{table}
\caption{The cuts used to extract the final radiative and $\pi^0$ counts. (See text for details.)  }
\begin{center}
\begin{tabular}{lc}
Cut Used & (Applied) \\ \hline
(1) $|M(\pi_1^{+}\pi^-)-M_{K^0}|<0.01$ GeV& ($both$) \\
(2) $|M(\pi_2^{+}n)-M_{\Sigma^{+}}|<0.01$ GeV & ($both$) \\
(3) $E_{x}<0.24$ GeV&($both$)\\
(4) $|M_x(\pi_1^{+}\pi^-)-M_{\Sigma^{*+}}|<0.03$ GeV& ($both$)\\
(5) $P^a_{\gamma}(\chi^2)<0.01\%$ &($\pi^0$)\\
(6) $P^b_{\pi}(\chi^2)>10\%$ &($\pi^0$)\\
(7) $P^a_{\pi}(\chi^2)<0.01\%$ &($\gamma$)\\
(8) $P^b_{\gamma}(\chi^2)>10\%$ &($\gamma$)\\\hline
\end{tabular}
\end{center}
\label{cuts_list}
\end{table}

\begin{figure}
\epsfig{file=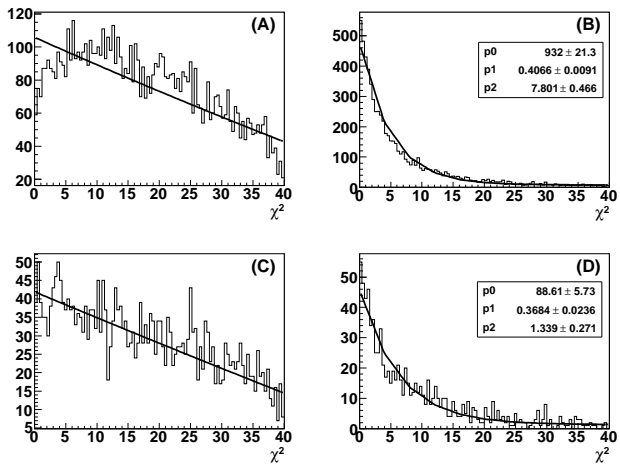,width=\columnwidth}
\caption{Plot (A) shows the $\chi^2$ distribution for the Monte Carlo channel 
$\gamma p \to K^0 \Sigma^{*+} \to \pi^+\pi^-\pi^+n \pi^0$ under the radiative hypothesis,
displaying a highly distorted $\chi^2$ distribution.  Plot (B) shows the fit results for $\gamma p \to K^0 \Sigma^{*+} \to \pi^+\pi^-\pi^+n \gamma$,
displaying a reasonable $\chi^2$ distribution with the $ideal$ $P_{1}$.
The radiative hypothesis kinematic fit $\chi^2$ distribution and distribution fit to data before any
$P_{\pi}^{a}(\chi^{2})$ cut is applied is shown in (C) and the same distribution after a $P_{\pi}^a(\chi^2)<0.01\%$ cut is
applied is shown in (D).}
\label{fig10-Chi2}
\end{figure}

The two-step kinematic fitting procedure was used to isolate the radiative
signal from the $\pi^{0}$ channel.  In this procedure two separate kinematic fits were performed, one with
zero missing mass for the $\gamma$ and the other with the missing mass of the $\pi^{0}$.  
The fit function in Eq. \ref{fittingf} was used to fit the $\chi^2$ distributions to determine the resulting quality of candidates present in the fit.
The parameter $P_{1}$ was used to measure how close the distribution in the histogram was to the $ideal$ theoretical $\chi^{2}$ distribution for two degrees of freedom.
The pure radiative decay Monte Carlo was used to determine the value of the expected $P_{1}$ parameter.
Fig. \ref{fig10-Chi2} (A) shows the $\chi^2$ distribution for a kinematic fit of the Monte Carlo channel $\gamma p \to K^0
\Sigma^{*+}\to\pi^+\pi^-\pi^+n\pi^0$ under the radiative hypothesis, displaying a highly distorted $\chi^2$ 
distribution. Fig. \ref{fig10-Chi2} (B) shows the $\chi^2$ distribution from the same kinematic fit under a
radiative hypothesis of the Monte Carlo channel $\gamma p \to K^0 \Sigma^{*+} \to \pi^+\pi^-\pi^+n
\gamma$, a fit using Eq. \ref{fittingf} after all the kinematic cuts up to (4) in Table \ref{cuts_list}.
The parameter $P_{1}$ was used as the $ideal$ value to expect.   After obtaining the
expected $P_1$, a kinematic fit to the data was performed using both hypotheses.

The first confidence level cut $P_{\pi}^{a}(\chi^{2})$ was used to filter out the larger $\pi^{0}$ 
signal from the radiative signal by using a kinematic fit to 
a $\Sigma^+ (\pi^{0})$ hypothesis and taking only the {\it low} confidence level candidates.  This
confidence level cut was checked and made more restrictive until $P_1$ from the data matched the expected value
from Monte Carlo (within the projected error bars).  Fig. \ref{fig10-Chi2} (C) shows the $\chi^2$ distribution and
fit before any $P_{\pi}^{a}(\chi^{2})$ cut was applied and Fig. \ref{fig10-Chi2} (D) shows the distribution after a $P_{\pi}^a(\chi^2)<0.01\%$ cut was applied.
The final confidence level cut $P_{\gamma}^{b}(\chi^{2})$ from the kinematic fit to a $\Sigma^+(\gamma)$ hypothesis was used on the remaining candidates.  Only the {\it high} confidence level candidates were preserved. 

\begin{figure}
\epsfig{file=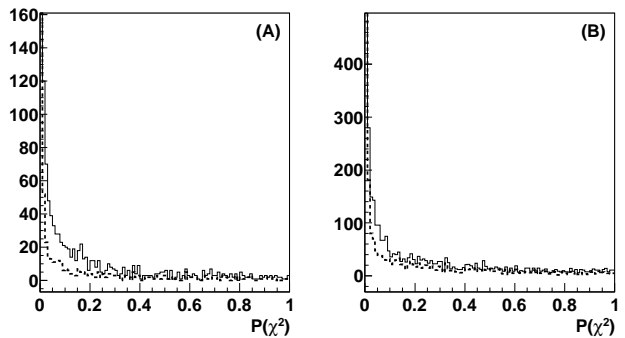,width=\columnwidth}
\caption{(A) The confidence level distribution for a fit to the data under the radiative hypothesis before the $P^a_{\pi}(\chi^2)<0.01\%$ cut (line) and after (dotted line).  (B) The confidence level distribution for a fit to the data under the missing $\pi^0$ hypothesis before the $P^a_{\gamma}(\chi^2)<0.01\%$ cut (line) and after (dotted line).}
\label{piCon}
\end{figure}

Note that the yields for the $\Sigma^+\pi^0$ decay will be reduced for a lower value of $P_\pi^a(\chi^2)$, 
which is desirable for extracting the radiative decay signal. On 
the other hand, this cut cannot be made arbitrarily small, since it 
reduces the statistics ($i.e.$, increases the statistical uncertainty). 
Similarly, the $\Sigma^+\gamma$ signal will be purified by a 
higher cut on $P_\gamma^b(\chi^2)$, but again the higher the cut, 
the lower the statistics.  The Monte Carlo was used to 
examine the acceptance of these cuts for various branching ratios 
($\Sigma^+ \gamma / \Sigma^+ \pi^0$), which is discussed in the next section.  Ultimately, the branching 
ratio extracted from the data should not depend on the cut points 
chosen (assuming the Monte Carlo gives accurate cut acceptances). 
The Monte Carlo was then used to optimize 
the trade-off between statistical uncertainty and systematic uncertainty (due 
to the choice of confidence level cuts based on a more quantitative analysis of $P_1$).
The cut value of $P^a(\chi^2)<0.01\%$ showed consistent optimization with $P^b(\chi^2)>10\%$ for
this topology and range of statistics as listed in Table \ref{cuts_list}.
Details of the optimization method of the confidence level cuts
using the Monte Carlo are described in Ref. \cite{keller3}.  The confidence level distribution under the
radiative hypothesis before and after the $P^a_{\pi}(\chi^2)<0.01\%$ cut is shown in Fig. \ref{piCon}(A).  Likewise, the confidence level distribution for a fit to the data under the missing $\pi^0$ hypothesis before and after the
$P^a_{\gamma}(\chi^2)<0.01\%$ cut is shown in Fig. \ref{piCon}(B).

\begin{figure}
\epsfig{file=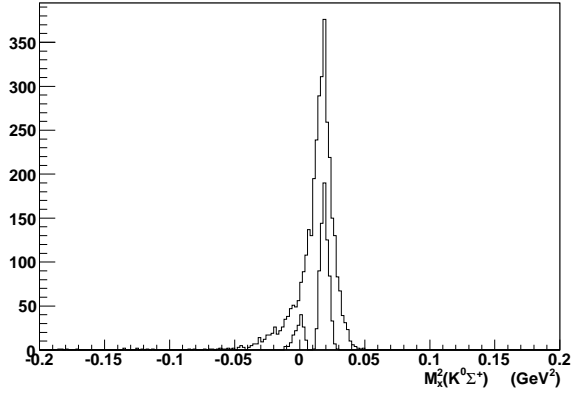,width=\columnwidth}
\caption{The missing mass off of all detected particles.  The plot shows the final radiative candidates
at zero missing mass after the $P^a_{\pi}(\chi^2)<0.01\%$ and $P^b_{\gamma}(\chi^2)>10\%$ cuts.  Also
shown are the final $\pi^0$ candidates after the $P^a_{\gamma}(\chi^2)<0.01\%$ and $P^b_{\pi}(\chi^2)>10\%$ 
cuts.}
\label{money}
\end{figure}

The same cuts determined to effectively isolate the radiative signal, $P^a_{\pi}(\chi^2)<0.01\%$ with
$P^b_{\gamma}(\chi^2)>10\%$, were used to isolate the $\pi^0$ channel used for normalization of the ratio,
such that $P^a_{\gamma}(\chi^2)<0.01\%$ is used with $P^b_{\pi}(\chi^2)>10\%$.  This can be denoted as
$P^a_{\pi}(\chi^2)=P^a_{\gamma}(\chi^2)<0.01\%$ and $P^b_{\gamma}(\chi^2)=P^b_{\pi}(\chi^2)>10\%$.  The
final missing mass squared distribution after all cuts is shown in Fig. \ref{money} before the two sets of confidence level cuts and after.    

The acceptances were found for each contributing channel and are listed in Table \ref{acc3}.  Each of the
channels in Table \ref{acc3} was generated with enough statistics so that the statistical uncertainty
would not contribute in the final ratio calculation.  The value of the acceptance for the
$\pi^0$ ($\gamma$) hypothesis is listed under the $A_{\pi}$ ($A_{\gamma}$) column.  The uncertainty
is statistical only.

\begin{table}
\caption{Acceptances (in units of $10^{-3}$) for the channels that survive all cuts.  All of the cuts used to obtain the acceptance values
are listed in Table \ref{cuts_list}. The uncertainties are statistical only.  The two columns
contain the acceptance for each hypothesis $A_{\pi}$, $A_{\gamma}$.}
\begin{center}
\begin{tabular}{lccc}
Reaction & $A_\pi$ & $A_\gamma$ &  \\ \hline
$K^{0}\Sigma^{*+}\rightarrow K^{0}\Sigma^{+}\gamma$
       & 0.0644$\pm$0.0040 & 0.6244$\pm$0.0125 & \\
$K^{0}\Sigma^{*+}\rightarrow K^{0}\Sigma^{+}\pi^{0}$
       & 0.6502$\pm$0.0128 & 0.0591$\pm$0.0038 & \\
$K^{*0}\Sigma^{+}\rightarrow K^{0}\Sigma^{+}\gamma$ 
       & 0.0018$\pm$0.0005 & 0.0186$\pm$0.0021 & \\
$K^{*0}\Sigma^{+}\rightarrow K^{0}\Sigma^{+}\pi^{0}$
       & 0.0231$\pm$0.0023 & 0.0050$\pm$0.0011 & \\
$\omega \Delta^+\rightarrow \pi^+\pi^-\pi^0 n \pi^{+} $ 
       & 0.0003$\pm$0.0001 & 0.0000$\pm$0.0000 & \\
$\omega \Delta^+\rightarrow \pi^+\pi^-\gamma n \pi^{+} $
       & 0.0000$\pm$0.0000 & 0.0002$\pm$0.0000 & \\     
       \hline
\end{tabular}
\end{center}
\label{acc3}
\end{table}

The acceptance values indicate that contributions from the $K^{*0}$ and $\omega$ channels will
be subtracted out directly.  All other background channels not listed were zero.  As mentioned, all
background contributions to the ratio are relatively small, but care is taken to accurately consider
each contribution.  The levels of these contributions depend the placement of the confidence level
cuts previously mentioned.  

\begin{figure}
\epsfig{file=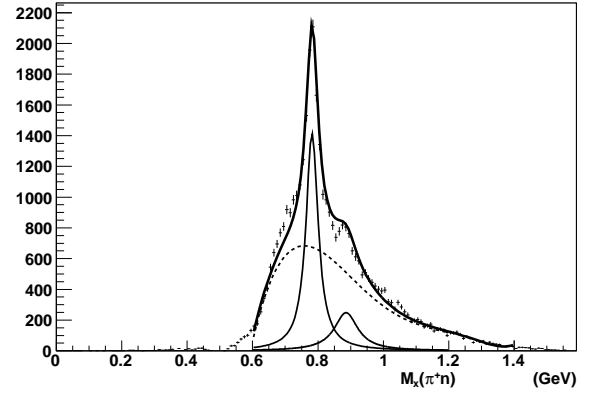,width=\columnwidth}
\caption{The missing mass off the $\pi_2^+$-$n$ combination with the $|M(\pi_1^{+}\pi^-)-M_{K^0}|<0.01$ GeV, 
$|M(\pi_2^{+}n)-M_{\Sigma^{+}}|<0.01$ GeV, and $|M^2_x-M^2_{\pi^{0}}|<0.0175$ GeV$^2$ cuts.  The resulting
$K^{*0}$ and $\omega$ peaks are fit with Breit-Wigner line shapes, while the background is fit with a polynomial function.}
\label{background}
\end{figure}

To obtain an estimate of the amount of leakage into the $\Sigma^+\gamma$ and $\Sigma^+\pi^0$ signals, some cuts were removed to obtain a fit on the channels of interest.  Only the $|M(\pi_1^{+}\pi^-)-M_{K^0}|<0.01$ GeV 
and the $|M(\pi_2^{+}n)-M_{\Sigma^{+}}|<0.01$ GeV cuts from Table \ref{cuts_list} were used with an additional cut on the missing mass squared of all detected particles around the $\pi^0$ mass of $|M^2_x-M^2_{\pi^{0}}|<0.0175$ GeV$^2$.  The missing mass off
the $\pi_2^{+}n$ combination was then checked.  The resulting
$K^{*0}$ and $\omega$ peaks were fit with a relativistic Breit-Wigner while the background was fit with a polynomial function, as shown in Fig. \ref{background}. 

The total number of $K^{*0}$ events present in the data set using the less restricted set
of cuts just described can be expressed as 
\begin{eqnarray}
N(K^{*0})=\nonumber\\
\frac{n(K^{*0}\to\pi^+\pi^-\pi^0)}{R(K^{*0}\to\pi^+\pi^-\pi^{0})A^{K^{*0}}(K^{*0}\to\pi^+\pi^-\pi^{0})}=\nonumber\\
3.639\times10^6\pm6.6\times10^4,
\label{eqK1}
\end{eqnarray}
where $n(K^{*0}\to\pi^+\pi^-\pi^0)=3019\pm55$ is the estimated number of $\gamma p \to K^{*0} \Sigma^+ \to \pi^+\pi^-\pi^0 \pi^+n$
events found through the integrated fit to the $K^{*0}$ in Fig. \ref{background}.  $R(K^{*0}\to\pi^+\pi^-\pi^{0})=1/3$
is the probability for the decay and $A^{K^{*0}}(K^{*0}\to\pi^+\pi^-\pi^{0})$ is the acceptance of the $K^{*0}\to\pi^+\pi^-\pi^{0}$ channel
found by observing how many thrown events survive the three cuts used to obtain the $K^{*0}$ sample.  The value $N(K^{*0})$ can then be used to obtain an estimate of the $K^{*0}$ contribution from any set
of cuts, given an accurate acceptance for the new cuts.
The number of $K^{*0}$ events that would be present in the analysis outlined in Table \ref{cuts_list} under the $\pi^0$ hypothesis can be expressed as
\begin{eqnarray}
N_{\pi}(K^{*0}\to\pi^+\pi^-\pi^0)=\nonumber\\
\frac{A_{\pi}(K^{*0}\to\pi^+\pi^-\pi^0)n(K^{*0}\to\pi^+\pi^-\pi^0)}{R(K^{*0}\to\pi^+\pi^-\pi^{0})A^{K^{*0}}(K^{*0}\to\pi^+\pi^-\pi^{0})},
\label{eqK2}
\end{eqnarray}
where $A^{K^{*0}}(K^{*0}\to\pi^+\pi^-\pi^{0})$ is the acceptance for the $\gamma p \to K^{*0} \Sigma^+ \to \pi^+\pi^-\pi^0 \pi^+n$
channel under the $\Sigma^+(\pi^0)$ hypothesis.  Likewise for the $\Sigma^+(\gamma)$ hypothesis interchanging $A_{\pi}(K^{*0}\to\pi^+\pi^-\pi^0)$ with $A_{\gamma}(K^{*0}\to\pi^+\pi^-\pi^0)$
to obtain $N_{\gamma}(K^{*0}\to\pi^+\pi^-\pi^0)$.  The $K^{*0}$ radiative decay can also have a contribution under the $\Sigma^+(\pi^0)$ hypothesis,
\begin{eqnarray}
N_{\pi}(K^{*0}\to\pi^+\pi^-\gamma)=\nonumber\\R(K^{*0}\to\pi^+\pi^-\gamma)A_{\pi}(K^{*0}\to\pi^+\pi^-\pi^0)N(K^{*0}),
\label{eqK3}
\end{eqnarray}
or under the $\Sigma^+(\gamma)$ hypothesis,
\begin{eqnarray}
N_{\gamma}(K^{*0}\to\pi^+\pi^-\gamma)=\nonumber\\R(K^{*0}\to\pi^+\pi^-\gamma)A_{\gamma}(K^{*0}\to\pi^+\pi^-\pi^0)N(K^{*0}),
\label{eqK4}
\end{eqnarray}
where $R(K^{*0}\to\pi^+\pi^-\gamma)=2.39\pm0.21\times10^{-3}$ \cite{pdg}.

In the case of the $\omega$ contributions, no distinction is made between the $\gamma p \to\omega \Delta^+$ and
$\gamma p \to\omega n\pi^+$ channels.  The Monte Carlo used in the background estimate for the $\omega$ is the $\gamma p \to\omega \Delta^+$
channel only because there is a slightly larger acceptance for this channel.  By using the channel of greatest acceptance an over-estimate is expected.
The total number of events from $\omega \Delta^+ \rightarrow \pi^+\pi^-\pi^0 n \pi^{+}$ present is estimated as
\begin{eqnarray}
N(\omega)=\nonumber\\\frac{n(\omega\to\pi^+\pi^-\pi^{0})}{R(\omega\to\pi^+\pi^-\pi^{0})A^{\omega}(\omega\to\pi^+\pi^-\pi^{0})}=\nonumber\\
5.296\times10^7\pm4.7\times10^5.
\label{eqOM}
\end{eqnarray}
Here $n(\omega\to\pi^+\pi^-\pi^{0})=11120\pm106$ is the estimate from integrating the $\omega$ fit in Fig. \ref{background}, $R(\omega\to\pi^+\pi^-\pi^0)=89.2\pm0.7\%$ is the branching ratio of the $\omega$ decay to $\pi^+\pi^-\pi^0$ \cite{pdg} and
$A^{\omega}(\omega\to\pi^+\pi^-\pi^0)$ is the probability that this decay channel will be observed after the three cuts used to obtain the fit to the $\omega$ peak.
An estimate of the number of counts under the $\Sigma^+(\pi^{0})$ hypothesis coming from the $\omega$ is obtained using Eq. \ref{eqOM} as
\begin{eqnarray}
N_{\pi}(\omega\to\pi^+\pi^-\pi^0)=\nonumber\\A_{\pi}(\omega\to\pi^+\pi^-\pi^0)R(\omega\rightarrow \pi^+\pi^-\pi^0)N(\omega),
\label{eqOM2}
\end{eqnarray}
where $A_{\pi}(\omega\to\pi^+\pi^-\pi^{0})$ is the acceptance for the $\gamma p \to \omega \Delta^+ \to \pi^+\pi^-\pi^0 \pi^+n$
channel under the $\Sigma^+(\pi^0)$ hypothesis. 

It is possible to express all other associated $\omega$ corrections to be obtained given the value of $N_{\omega}$, 
along with the acceptance terms for that particular channel.  The corrections for the $\gamma$ and $\pi$ channels, respectively,
are written as 
\begin{eqnarray}
N_{\gamma,\pi}(\omega\rightarrow \pi^+\pi^-\pi^0\pi^0)&=&\nonumber\\
A_{\gamma,\pi}(\omega\rightarrow\pi^+\pi^-\pi^0\pi^0)R(\omega\rightarrow \pi^+\pi^-\pi^0\pi^0)N(\omega),\nonumber\\
N_{\gamma,\pi}(\omega\rightarrow\pi^+\pi^-\gamma)&=&\nonumber\\
A_{\gamma,\pi}(\omega\rightarrow\pi^+\pi^-\gamma)R(\omega \to\pi^+\pi^-\gamma)N(\omega),\nonumber\\
N_{\gamma,\pi}(\omega\rightarrow\pi^+\pi^-\pi^0)&=&\nonumber\\
A_{\gamma,\pi}(\omega\rightarrow\pi^+\pi^-\pi^0)R(\omega \rightarrow \pi^+\pi^-\pi^0)N(\omega).
\end{eqnarray}
where $R$ is used for the corresponding branching ratio or upper limit in each case, for example, $R(\omega \to\pi^+\pi^-\gamma)$ is the branching ratio for the radiative decay of the $\omega$
with a value less than $3.6 \times 10^{-3}$.  The value
of $R(\omega\rightarrow\pi^+\pi^-\pi^0\pi^0)$ is listed at less than $2\times10^{-4}$ \cite{pdg}.  All results from background
contributions are tabulated in Table \ref{bg}.  All non-listed background is considered negligible.

\begin{table*}
\caption{The contributions for all of the background channels taken into consideration.  The uncertainties are statistical only.  The two columns
contain the contributions estimated for each hypothesis $N_{\gamma}$, $N_{\pi}$.}
\begin{center}
\begin{tabular}{lccc}
Reaction & $N_\pi$ & $N_\gamma$ &  \\ \hline
$K^{*0}\Sigma^{+}\rightarrow K^{0}\Sigma^{+}\pi^0$ & 28.02$\pm$2.84 & 6.02$\pm$1.34 & \\
$K^{*0}\Sigma^{+}\rightarrow K^{0}\Sigma^{+}\gamma$ & 0.0157$\pm$0.0046 & 0.162$\pm$0.023 & \\
$\omega \Delta^+\rightarrow \pi^+\pi^-\pi^0 n \pi^{+}$ &$ 11.81\pm2.37 $&$ 0.0120\pm0.0024$ & \\
$\omega \Delta^+\rightarrow \pi^+\pi^-\gamma n \pi^{+}$ &$ 0.0400\pm0.0057$&$ 2.1\times10^{-5}\pm3.0\times10^{-6}$ & \\
$\omega \Delta^+\rightarrow \pi^+\pi^-\pi^0 \pi^0 n \pi^{+}$ &$ 1.1\times10^{-4}\pm2.0\times10^{-5} $&0.0 & \\
       \hline
\end{tabular}
\end{center}
\label{bg}
\end{table*}

\subsection{Final yields}
\label{results1}

To calculate the ratio of the EM decay to the strong decay, Eq. \ref{eq_newR} is employed.  All terms that take into account
any channel other than the $\pi^0$ and radiative signals are for the time being ignored.  The
acceptance values are taken from Table \ref{acc3}.  The raw values obtained out of the final kinematic fit are $n_{\gamma}= 148$ and $n_\pi = 682$, as seen in Fig. \ref{money}, with statistical uncertainties taken as the square root of $n$ in each case.  After accounting for the backgrounds listed in Table \ref{bg}, the corrected counts are $\Delta n_{\gamma}= 135.81\pm11.99$ and $\Delta n_\pi = 642.11\pm26.38$.

The ratio of the $K^{0}\Sigma^{*+}\rightarrow K^{0}\Sigma^{+}\gamma$ channel to the $K^{0}\Sigma^{*+}\rightarrow
K^{0}\Sigma^{+}\pi^0$ channel is then,
\begin{eqnarray}
R^{\Sigma^{*+}\to \Sigma^+\gamma}_{\Sigma^{*+}\to \Sigma^+\pi^0} &=&\frac{\Delta n_{\gamma}A_{\pi}(\Sigma\pi)-\Delta n_{\pi}A_{\gamma}(\Sigma\pi)}{\Delta n_{\pi}A_{\gamma}(\Sigma\gamma)-\Delta n_{\gamma}A_{\pi}(\Sigma
\gamma)}\nonumber\\&=&11.95\pm2.21\%.
\end{eqnarray}

The raw counts for the radiative and $\pi^{0}$ extraction were obtained using $P^a_{\gamma}=P^a_{\pi}<0.01\%$ with
the final confidence level cut $P^b_{\gamma}=P^b_{\pi}>10\%$ as mentioned in the
cuts from Table \ref{cuts_list}.  

Only the statistical uncertainty is quoted.  To determine how reliable the ratio is a set of
systematic studies is required along with a study of the variation
in the ratio based on the choice of confidence level cuts.  
This variation and all other systematic studies are considered in the next section.

\section{Systematic Studies}
\label{sysStud2}
The value of each of the nominal cuts was varied to study the effect on the final background corrected ratio.
For each variation the new acceptance terms in Equation \ref{eq_newR} were
recalculated with the corresponding Monte Carlo.  Each major systematic uncertainty
contribution is numbered as it is discussed and listed in Table \ref{systematicX}, which 
contains a summary of all systematic uncertainties. 

Several $\Delta \beta$ cut variations were checked starting with $|\Delta \beta| < 0.02$ for all charged particles, leading
to a ratio of 11.98$\pm$2.22$\%$.  There was also a check at $|\Delta \beta| < 0.1$ that gave a ratio of
11.74$\pm$2.17\%.  The $|\Delta \beta|$ cut selected uses a $\pm 1$ ns timing cut, while keeping $|\Delta
\beta| < 0.035$ for all pions.  This variation is presented in Table \ref{systematicX} as number (1).

To estimate the systematic effects from the Monte Carlo, such as the uncertainty in correctly simulating the data,
a comparison was made with the cross section of $\gamma p \to K^0 \Sigma^{*+}$ from Monte Carlo and data.  The ratio was obtained using the 
acceptance corections based on a Monte Carlo with a zero $t$-slope in the generator to get 
$R=11.70\pm$2.21$\%$.  This value deviated from the ratio obtained by $\sim2\%$.  The uncertainty from Monte Carlo was then estimated to be $\sim2\%$ in either direction.  This finding leads to a high value in the ratio of $R=12.19\pm$2.21$\%$ and a lower value in the ratio of $R=11.71\pm$2.21$\%$, which is listed as number (2) in Table \ref{systematicX}.

The missing energy cut
removes a large amount of background that slips in under the
$\Sigma^{*+}$.  The final cut of $E_x<0.24$ GeV was chosen based on the maximization of signal counts
(decreased statistical uncertainty).  The systematic contribution for the $E_x$ cut was studied by varying the cut in a reasonable range
about the nominal value.  This ratio was reasonably stable up
until $E_x<0.28$ GeV, at which point the $\pi^0$ and $\gamma$ signals become overwhelmed with background.  Cuts in $E_x$ too low in energy tend to distort the ratio.
Based on this study a high ($11.95\pm$2.21$\%$) and low ($11.71\pm$2.20$\%$) value were assigned to the associated systematic uncertainties.  These contributions are listed in Table \ref{systematicX} as number (3).

The background counts that contribute to the ratio assume a branching ratio for $\omega\to\pi^+\pi^-\gamma$ to be at the top of the upper limit at
$3.6\times10^{-3}$, as well as for $\omega\to\pi^+\pi^-\pi^0\pi^0$ at
$2\times10^{-4}$ \cite{pdg}.  The variation in uncertainty of the branching ratio of
either of these channels is not of large enough order to make a notable effect on the ratio.
The total contribution from background in the $\pi^0$ hypothesis was $39.89\pm3.70$ and for the $\gamma$ hypothesis was $6.19\pm1.34$.  To check for the largest deviations under these uncertainties, a contribution of
43.59 counts from the $\pi^0$ hypothesis was used with 4.85 counts for the $\gamma$ hypothesis, leading to a ratio of $12.37\pm2.24\%$.  The opposite extreme was also used to obtain 36.19 counts from the $\pi^0$ hypothesis and 7.5 counts for the $\gamma$ hypothesis, leading to a ratio $11.54\pm2.19\%$.
This finding is listed in Table \ref{systematicX} as number (4). 

The kinematic fits used to control the leakage of the $K^{*0}$ and $\omega$ have an associated confidence level cut that was also tested.  The cut in each case was selected based on the reduction of the
specific background channel, while maximizing the signal counts from the $\gamma$ and $\pi^0$.  A check was done over a large range of confidence level cuts for each case to test the variation in the ratio.  New
background contributions were found for each cut along with new acceptance terms.  The ratio was then recalculated and tabulated for the $K^{*0}$ ($\omega$) removal in Table \ref{probbg1} (Table \ref{probbg2}).  These contributions are listed as (5) and (6), respectively, in Table \ref{systematicX}. 

\begin{table}
\begin{center}
\begin{tabular}{lccc}
$P_\gamma(\chi^2)=P_\pi(\chi^2)$ & $R$\\ \hline
0.100 &  12.04$\pm$2.20\% \\
0.050 &  11.79$\pm$2.20\% \\
0.010 &  11.95$\pm$2.21\% \\   
0.005 &  11.92$\pm$2.22\% \\
0.001 &  11.90$\pm$2.21\% \\ \hline
\end{tabular}
\end{center}
\caption{The values used in the systematic variation for the calculated ratio with respect to 
the confidence level cut from the fit to $\gamma p \to \pi^+ n K^{*0}\to \pi^+ n\pi^+\pi^-(\pi^0)$ and
$\gamma p \to \pi^+ n K^{*0} \to \pi^+ n \pi^+\pi^-(\gamma)$.  The confidence level cuts were set equal to each other such that $P_\gamma(\chi^2)=P_\pi(\chi^2)$.}
\label{probbg1}
\end{table}

\begin{table}
\begin{center}
\begin{tabular}{lccc}
$P_\gamma(\chi^2)=P_\pi(\chi^2)$ & $R$\\ \hline
0.100 &  11.99$\pm$2.20\% \\
0.050 &  12.01$\pm$2.21\% \\
0.010 &  11.95$\pm$2.21\% \\
0.005 &  11.80$\pm$2.22\% \\
0.001 &  11.77$\pm$2.21\% \\ \hline
\end{tabular}
\end{center}
\caption{The values used in the systematic variation for the calculated ratio with respect to 
the confidence level cut from the fit to $\gamma p \to \pi^+ n \omega\to \pi^+ n\pi^+\pi^-(\pi^0)$ and
$\gamma p \to \pi^+ n \omega\to \pi^+ n \pi^+\pi^-(\gamma)$.  The confidence level cuts were set equal to each other such that $P_\gamma(\chi^2)=P_\pi(\chi^2)$.}
\label{probbg2}
\end{table}

The optimum set of confidence level cuts used to extract the final yields has a range of validity in which a study of the ratio variation is appropriate.  The
range of validity is found by using the fractional deviation in the ratio $\delta R$, and requiring it to be less than or equal to the fractional uncertainty
due to statistics.  $\delta R$ is defined as the difference in the generated ratio and recovered ratio in the Monte Carlo study.  
The set of optimum cuts occurs at different values for a 
given mixture of Monte Carlo $\Sigma^+ \pi^0$ and $\Sigma^+ \gamma$ 
events.  Using the $g11a$ data, the Monte Carlo was tuned to have 
approximately the same ratio and the same statistics as the real data, 
and the optimum cuts were thereby determined quantitatively.
This set of cuts used in the systematic studies,
along with the values of $\delta R$, are given 
in Table \ref{opti} for isolation of the radiative decay and Table \ref{opt2} for the $\pi^0$ isolation.
Note that the value of $\delta R$ is {\it not} 
the systematic uncertainty in $R^{\Sigma^+\gamma}_{\Sigma^+\pi}$. 
Rather, the systematic uncertainty comes from the data ratio values in the right-most column in each table. The systematic uncertainty is based on the variation in the extracted ratio of Eq. (\ref{R}) for a set
of cuts determined to give the minimal deviation in the Monte Carlo recovered ratio for the valid range of statistics.

\begin{table}
\begin{center}
\begin{tabular}{lccc}
$P^a_{\pi}(\%)$ & $P^b_{\gamma}(\%)$ & $\delta$$R$ & $R$\\ \hline
0.050 & 17 & 0.140 & 11.68$\pm$2.22\%\\
0.005 & 15 & 0.090 & 11.88$\pm$2.29\%\\   
0.010 & 10 & 0.089 & 11.95$\pm$2.21\%\\
0.075 & 20 & 0.092 & 11.41$\pm$2.27\%\\ \hline
\end{tabular}
\end{center}
\caption{Optimization points for each $P^a_{\pi}$ and $P^b_{\gamma}$ for the $\Sigma^+\gamma$ channel.}
\label{opti}
\end{table}
\begin{table}
\begin{center}
\begin{tabular}{lccc}
$P^a_{\gamma}(\%)$ & $P^b_{\pi}(\%)$ & $\delta$$R$ & $R$\\ \hline
0.050 & 17 & 0.187 & 11.35$\pm$2.23\%\\
0.005 & 15 & 0.109 & 12.13$\pm$2.31\%\\   
0.010 & 10 & 0.099 & 11.95$\pm$2.21\%\\
0.075 & 20 & 0.195 & 11.19$\pm$2.30\%\\ \hline
\end{tabular}
\end{center}
\caption{Optimization points for each $P^a_{\gamma}$ and $P^b_{\pi}$ for the $\Sigma^+\pi^0$ channel.}
\label{opt2}
\end{table}
The systematic dependence on the choice of the $P(\chi^2)$ cuts in both the radiative and
$\pi^0$ hypotheses comes from Table \ref{opti} and Table \ref{opt2}.  The ratio was recalculated from the resulting raw counts in each case with the new acceptance terms for the set of cuts obtained from the optimal range.  The set of cuts was tested for both the radiative and $\pi^0$ hypotheses separately.  In each variation under the $\pi^0$ ($\gamma$) hypothesis, the opposing cut for the $\gamma$ ($\pi^0$) hypothesis was not changed.  The highest and lowest value from each study were used as the contributions
to the systematic uncertainty listed as (7) and (8) in Table \ref{systematicX}.

The cut on the $\pi_1^+$-$\pi^-$ invariant mass was used to minimize the events that are not
associated with a $K^0$.  A range of cuts on $|M(\pi_1^+\pi^-)-M_{K^0}|$ was used to study this effect.  Only
the stable region was
used in the final determination of the range of variation for this cut with a high value of 11.95$\pm$2.21\% and a low value of 11.72$\pm$2.21\%.  These are listed in line (9) of Table \ref{systematicX}.

The cut on the $\pi_2^+$-$n$ invariant mass regulates the candidates going into the final
set of kinematic fits so there is some sensitivity to any events that do not come from the $\Sigma^+$.
A range of cuts on $|M(\pi_2^+n)-M_{\Sigma^+}|$ was used to study this effect.  Only the stable region was
used in the final determination of the range of variation for this cut with a high value of 11.95$\pm$2.21\% and a low value of 11.34$\pm$2.85\%.  These are listed in line (10) of Table \ref{systematicX}.

Table \ref{systematicX} shows a summary of the systematic studies and the highest and lowest value of the
ratio based on the variations mentioned for each type of uncertainty.
\begin{table*}
\caption{Sources of systematic variation in the ratio (in \%) showing the contributions to the systematic uncertainties from changes to the event selection 
values, along with the low and high value used.}
\begin{center}
\begin{tabular}{lccccc}
Source & Low Value & Low Contribution & High Value & High Contribution \\ \hline
(1) $\Delta\beta$ &11.74$\pm$2.17 & -0.21 & 11.98$\pm$2.22 & +0.03   \\
(2) MC-match &11.71$\pm$2.21 & -0.24 & 12.19$\pm$2.21 & +0.24  \\
(3) $E_{x}$ &11.71$\pm$2.20 & -0.24 & $11.95\pm$2.21 & +0.00  \\
(4) BG uncertainty &11.54$\pm$2.19 & -0.41 &12.37$\pm$2.24  & +0.42  \\
(5) $K^{*0}$-CL &11.79$\pm$2.20  & -0.16 &12.04$\pm$2.20  & +0.09 \\
(6) $\omega$-CL &11.77$\pm$2.21 & -0.18 & 12.01$\pm$2.21 & +0.06 \\
(7) $\gamma$-CL &11.41$\pm$2.27 & -0.54 & 11.95$\pm$2.21 & +0.00  \\
(8) $\pi^0$-CL &11.35$\pm$2.23 & -0.60 & 12.13$\pm$2.31 & +0.18  \\
(9) $K^0$ cut &11.72$\pm$2.21 & -0.23 & 11.95$\pm$2.21 & +0.00  \\
(10) $\Sigma^+$ cut & 11.34$\pm$2.85 & -0.61 & 11.95$\pm$2.21 & +0.00  \\ \hline
Total Uncertainty &    & -1.21	&	&  +0.53 \\ \hline
\end{tabular}
\end{center}
\label{systematicX}
\end{table*}
To calculate the final systematic uncertainty, the difference in the ratio $R=11.95\%$ and the high value of the ratio for each case in 
Table \ref{systematicX} was added in quadrature to obtain a value for the uncertainty of 0.53\% greater than the ratio.  The lower systematic uncertainty bound
was based on the difference
between the ratio $R=11.95\%$ and the low value of the ratio for each case, resulting in a value of 1.21\% less than the ratio.
The final ratio reported is $11.95\pm2.21(stat)_{-1.21}^{+0.53}(sys)\%$.

\section{Overall Results}
The final result for the ratio of the $\Sigma^{*+}\to \Sigma^+\gamma$ to $\Sigma^{*+}\to \Sigma^+\pi^0$ with systematic uncertainties is 
\begin{eqnarray}
R^{\Sigma^{*+}\to \Sigma^+\gamma}_{\Sigma^{*+}\to \Sigma^+\pi^0} =\frac{n_{\gamma}A_{\pi}(\Sigma\pi)-n_{\pi}A_{\gamma}(\Sigma\pi)}{n_{\pi}A_{\gamma}(\Sigma\gamma)-n_{\gamma}A_{\pi}(\Sigma
\gamma)}=\nonumber\\11.95\pm2.21(stat)_{-1.21}^{+0.53}(sys)\%.
\end{eqnarray}
To calculate the EM decay partial width from the measured branching ratio, the full width of the $\Sigma^{*+}$ decay
is used, $\Gamma_{Full}=35.8\pm0.8$ MeV \cite{pdg}, with the branching ratio $R(\Sigma^{*+}\to\Sigma^{+}\pi^0)=5.85\pm0.75\%$.  The partial width
calculation including systematic uncertainties leads to,
\begin{eqnarray}
\Gamma_{\Sigma^{*+}\to \Sigma^{+}\gamma} =R^{\Sigma^{*+}\to \Sigma^+\gamma}_{\Sigma^{*+}\to \Sigma^+\pi^0}
R(\Sigma^{*}\to\Sigma^{+}\pi^0)\Gamma_{Full}=\nonumber\\250\pm56.9(stat)_{-41.2}^{+34.3}(sys)\text{ keV}.
\end{eqnarray}
To obtain the corresponding U-spin prediction, we first look at the prediction for the $\Delta^+ \to p \gamma$ partial width to the $\Sigma^{*+}\to \Sigma^+\gamma$ partial width.
\begin{eqnarray} 
\left\langle \Delta^{+}|p \gamma \right\rangle &=&\left\langle \frac{1}{2} -\frac{1}{2} |\frac{1}{2} -\frac{1}{2}~0~~0 \right\rangle = 1 \nonumber\\
\left\langle \Sigma^{*+}|\Sigma^+\gamma\right\rangle&=&\left\langle \frac{1}{2} +\frac{1}{2} |\frac{1}{2}
-\frac{1}{2}~0~~0 \right\rangle= 1,\nonumber
\end{eqnarray}
leading to a ratio of, 
\begin{eqnarray}
\frac{\Gamma(\Delta^{+} \to p\gamma)}{\Gamma(\Sigma^{*+}\to \Sigma^+\gamma)}=
\left(\frac{M_{p}}{M_{\Delta}}\right)\left(\frac{M_{\Sigma^+}}{M_{\Sigma^{*+}}}\right)^{-1}\left(
\frac{q_{p}}{q_{\Sigma^+}}\right)^3=2.638.\nonumber
\end{eqnarray}
The value for the center of mass momentum for the proton is $q_p=0.259$ GeV and for the $\Sigma^+$ is $q_\Sigma^+=0.180$ GeV, \cite{pdg}.

This implies that the U-spin prediction for the partial width of the electromagnetic decay using the
$660\pm60$ MeV width of the $\Delta^{+} \to p\gamma$ decay \cite{pdg} is,
\begin{eqnarray}
\frac{\Gamma(\Delta^{+} \to p\gamma)}{2.638} =250\pm23\text{ keV}.
\end{eqnarray}
A similar calculation can be done for to obtain a U-spin prediction for the $\Gamma_{\Sigma^{*0}\to\Lambda\gamma}$ partial decay width.
Table \ref{tab:final} shows the previous model predictions along with the U-spin prediction and the final 
results from this analysis in each case.  The statistical and systematic uncertainties are combined in the present partial widths.

\begin{table}[h]
\caption{Comparison of theoretical model predictions for the radiative decay widths
with the experimental result for $\Sigma^{*0}$ from Ref. \cite{kellpaper} and the
present result result for $\Sigma^{*+}$.}
\begin{center}
\begin{tabular}{lcc}  
Model &$\Sigma(1385)^0\to\Lambda\gamma$&$\Sigma^+(1385) \to \Sigma^+\gamma$\\
\hline\hline
NRQM \cite{DHK,Koniuk,kaxiras}	& 273	&  104	\\
RCQM \cite{warns}   		& 267	&  	\\
$\chi$CQM \cite{wagner} 		& 265	&  105	\\
MIT Bag \cite{kaxiras}  		& 152	&  117	\\
Soliton \cite{Schat} 		& 243	&  91	\\
Skyrme \cite{Abada,Haberichter}	& 157-209	&  47	\\
Algebraic model \cite{Bijker} 	& 221.3	&  140.7 \\
U-spin 		        & 423$\pm$38&250$\pm$23\\\hline
Results  		& 445$\pm$102 &250$\pm$70 \\
\hline\hline
\end{tabular}   
\end{center}
\label{tab:final}
\end{table}

The partial width in both cases is larger than any prediction listed except for the U-spin prediction.
The U-spin predictions for the $\Sigma^{*0}\to \Lambda \gamma$ and $\Sigma^{*+}\to \Sigma^+ \gamma$ partial widths are well validated by the
experimental result.  For these cases U-spin symmetry is confirmed within the experimental
uncertainties.
It is important to note that the U-spin prediction for the $\Sigma^{*+}$ EM decay partial width ignores the
effects of the interference of the isovector and isoscalar components of the photon.  If the isoscalar
component interfered destructively, the resulting prediction could indeed be much smaller.

The results in Ref. \cite{lee2} reveal that the meson cloud effect can contribute significantly ($\sim40\%$) to the overall
electromagnetic decay width of the $\Delta\to N\gamma$.  This puts the prediction from the model at about 
$80\%$ of the experimental measurement.  As stated previously it has not yet been
determined from a theoretical standpoint if the meson cloud effects contribute and if so
to what degree for the radiative decay of the $\Sigma^{*0}$ and $\Sigma^{*+}$.  This may be the reason for such
a difference in the predictions seen from experiment compared to the models listed in Table \ref{tab:final}.  Because the U-spin prediction for the $\Sigma^{*}$ EM decay width 
uses empirical information from the $\Delta$ EM decay, contributions from phenomena like the meson cloud effect should
be inherent.  The correspondence between the U-spin prediction and the experimental result strongly suggests that
the other models lack that corresponding effect.

Perhaps this work can prompt more encompassing 
calculations that are necessary to probe the structure of the baryon resonances and motivate
consideration of the meson cloud contributions for electromagnetic decay predictions.

\section{Acknowledgment}
The authors thank the staff of the Thomas
Jefferson National Accelerator Facility who made this experiment possible.
This work was supported in part by 
the Chilean Comisi\'on Nacional de Investigaci\'on Cient\'ifica y Tecnol\'ogica (CONICYT),
 the Italian Istituto Nazionale di Fisica Nucleare,
the French Centre National de la Recherche Scientifique,
the French Commissariat \`{a} l'Energie Atomique,
the U.S. Department of Energy,
the National Science Foundation,
the UK Science and Technology Facilities Council (STFC),
the Scottish Universities Physics Alliance (SUPA),
the United Kingdom's Science and Technology Facilities Council,
and the National Research Foundation of Korea.

The Southeastern Universities Research Association (SURA) operates the
Thomas Jefferson National Accelerator Facility for the United States
Department of Energy under contract DE-AC05-84ER40150.

\end{document}